\documentclass[sigconf]{nimeart}

\usepackage{subcaption}
\usepackage[dvipsnames]{xcolor}
\usepackage{array,multirow,graphicx}
\usepackage{booktabs}
\usepackage{csquotes}
\usepackage{balance}
\newcommand{\STAB}[1]{\begin{tabular}{@{}c@{}}#1\end{tabular}}

\AtBeginDocument{%
  }

\setcopyright{cc}
\copyrightyear{2026}
\acmYear{2026}
\acmDOI{}

\acmConference[NIME '26]{International Conference on New Interfaces for Musical Expression}{June 23--26,
  2026}{London, UK}
\acmISBN{}
\begin{document}


\title{\anon[Anonymous Name]{Latent Terrain}: Adapting Neural Audio Autoencoders as Design Materials in NIME}

\author{Shuoyang Jasper Zheng}
\orcid{0000-0002-5483-6028}
\affiliation{%
  \institution{Centre for Digital Music, Queen Mary University of London}
  \city{London}
  \country{UK}
}
\email{shuoyang.zheng@qmul.ac.uk}

\author{Keigo Yoshida}
\orcid{0009-0008-4518-0662}
\affiliation{%
  \institution{Graduate School of Media and Governance, Keio University}
  \city{Kanagawa}
  \country{Japan}}
\email{Keigoyoshida@keio.jp}

\author{Nico García-Peguinho}
\affiliation{%
  \institution{Centre for Digital Music, Queen Mary University of London}
  \city{London}
  \country{UK}
}
\email{n.a.garcia-peguinho@qmul.ac.uk}

\author{Jiatong Liu}
\orcid{0009-0008-3044-4442}
\affiliation{%
  \institution{Creative Computing Institute, University of the Arts London}
  \city{London}
  \country{UK}
}
\email{jiatong.liu37@outlook.com}

\author{Dan Hearn}
\affiliation{%
  \institution{Creative Computing Institute, University of the Arts London}
  \city{London}
  \country{UK}
}
\email{d.hearn@arts.ac.uk}

\author{Anna Xambó Sedó}
\orcid{0000-0003-2333-6941}
\affiliation{%
  \institution{Centre for Digital Music, Queen Mary University of London}
  \city{London}
  \country{UK}
}
\email{a.xambosedo@qmul.ac.uk}

\author{Nick Bryan-Kinns}
\orcid{0000-0002-1382-2914}
\affiliation{%
  \institution{Creative Computing Institute, University of the Arts London}
  \city{London}
  \country{UK}
}
\email{n.bryankinns@arts.ac.uk}

\renewcommand{\shortauthors}{Zheng et al.}

\begin{abstract}

Neural audio autoencoders, a deep learning method for sound synthesis, are increasingly popular in AI-enhanced NIMEs. It is timely to explore as a community how this technological opportunity has opened a domain of design in NIME. This paper focuses on one compelling technique of using autoencoders for sound synthesis: navigating their latent space as a generative sound space. We introduce \textit{\anon[Anonymous Name]{Latent Terrain}}, a Max/MSP tool package designed to tailor latent spaces into corpus-based sound spaces for NIMEs. We describe the rationale and development process of \anon[Anonymous Name]{Latent Terrain} to offer insights into the use of autoencoders in a material-oriented crafting space of musical interface design. We present an annotated portfolio resulting from a collaborative artistic exploration of \anon[Anonymous Name]{Latent Terrain} with four NIME makers, to showcase the design possibilities opened by autoencoders. We reflect on our practice-based account to discuss the challenges and opportunities of enabling neural audio autoencoders as design materials for AI-enhanced NIMEs.

\end{abstract}


\maketitle

\section{Introduction}
\label{sec_intro}

Neural audio autoencoders (or neural audio codecs), a deep learning method for sound synthesis \citep{engel_neural_2017, caillon_rave_2021}, have become increasingly popular in developing AI-enhanced sound and music applications \citep{evans_stable_2025, caspe_waveform_2025}. Research in Music Information Retrieval yields models that can effectively learn compact latent representations from audio for sound analysis and synthesis \citep{pasini_music2latent_2024}. In the field of NIME, such a unique affordance is enabling new ways of approaching musical interaction design \citep{yee-king_latent_2022, tahiroglu_latent_2024}. Neural audio autoencoders are increasingly engaged as interactive, exploratory, and expressive computational materials \citep{privato_magnetic_2023, scurto_soundwalking_2023, vigliensoni_steering_2023, nabi_embodied_2024, privato_stacco_2024, symons_perceptron_2024, khallaghi_squishysonics_2025}, expanding the ways of using and appropriating AI in musical interfaces. Given the multifaceted challenges and opportunities of neural audio autoencoders, it is timely to explore as a community how they have opened a domain of design \citep{gaver_annotated_2012} in NIME.

One approach to unpack this domain of design is through the lens of \textit{material-oriented musical interaction} \citep{mudd_material-oriented_2019} -- that is, acknowledging technologies' role in shaping creative actions, and treating them as the site for exploring musical activities \citep{mudd_material-oriented_2019, zheng_material_2022}. In particular, it draws attention to how makers craft and experiment with technologies to explore their use ``at play'' in interaction design \citep{scurto_prototyping_2021, caramiaux_explorers_2022}. NIME research has employed this material-oriented view to observe the emerging affordances of tools and technologies in musical activities \citep{zappi_hackable_2018, robson_thinking_2024}. 

However, we argue that artists and designers' hands-on engagement with neural audio autoencoders remains constrained for various reasons. First, technical AI research is typically positioned outside audio production and creative coding environments, offering limited means for NIME crafting \citep{caspe_waveform_2025, scurto_prototyping_2021}. Second, limited resources such as technical support and documentation raise the barrier to learning and using the underlying models and data \citep{bryan-kinns_xaixarts_2025, clarke_longevity_2025}. Last, the opaque and high-dimensional nature of model parameters \citep{bryan-kinns_exploring_2021} makes their sonic behaviours harder to understand than conventional NIME materials. In other words, the need for tools and resources that enable artist-led and practice-led engagement with autoencoders has emerged.

This paper focuses on one compelling technique of using autoencoders, the \textit{decoder-only} technique that involves overriding latent representations with user inputs to use the decoder as a sound synthesis model \cite{vigliensoni_steering_2023}. As a platform to explore nuanced control over AI generative models \cite{yee-king_latent_2022, tahiroglu_latent_2024}, the decoder-only technique allows musicians to ``navigate'' or ``walk'' in the latent space as a sound space for musical expressions \cite{scurto_soundwalking_2023, nabi_embodied_2024}. We contribute \textit{\anon[Anonymous Name]{Latent Terrain}},\footnote{Project homepage: \anon[Anonymous link]{\url{https://jasper-zheng.github.io/nn_terrain/}}} a Max/MSP package with a repository of resources for sound synthesis with neural audio autoencoders. \anon[Anonymous Name]{Latent Terrain} allows makers to tailor autoencoders into customised sound spaces that yield spectrally complex and varied musical phrases,\footnote{See Section 3.1.2 for details.} aiming to bridge autoencoders with broader NIME materials. We use this package to explore 
our research question: \textit{What are the challenges and opportunities of using neural audio autoencoders as design materials for NIME makers?} 


We offer an account of how we transform our technical experimentation of neural audio autoencoders into the Max/MSP package. Inspired by practice-led works that embrace artistic contextualisation of the artefact during its iteration \citep{gioti_composing_2022, armitage_tolvera_2024, shier_designing_2025}, we engage in a collaborative exploration with four artists and musicians to illustrate a design space of \anon[Anonymous Name]{Latent Terrain} in NIME. This leads to the creation of an annotated portfolio \cite{gaver_annotated_2012} that offers insight into the design thinking and experience behind four artistic pieces. Together, these technical practices and artistic practices allow us to reflect on the opportunities in NIME that have opened by neural audio autoencoders, and the challenges of adapting them as materials in a crafting space.

\section{Background}
\label{sec_background}

Here we introduce the technical background of autoencoders, the decoder-only technique, and the corpus-based method to build musical interfaces. We also introduce our perspective, the material-oriented exploration of neural audio autoencoders in Digital Musical Instrument (DMI) design.

\subsection{Audio Latent Space and Latent Trajectories}
\label{sec_trajectories}
There have been comprehensive reviews of autoencoders and latent spaces \cite{yee-king_latent_2022, tahiroglu_latent_2024}. Here we summarise their essential background. An audio latent space is a multi-dimensional vector space learned from a corpus of audio data \citep[p.~501]{goodfellow_deep_2016}, typically learned by an autoencoder. The components of an autoencoder include (i) an encoder that compresses audio waveforms into a sequence of vectors in the latent space, referred to as a \textbf{latent trajectory}, and (ii) a decoder that takes the latent trajectory to reconstruct audio waveforms, which aims to resemble the original waveforms. 

An emerging way of using autoencoders for sound synthesis in NIME design is a \textbf{decoder-only} approach that involves overriding the latent vectors with continuous user input to create synthetic latent trajectories \citep{weber_parameterized_2019, vigliensoni_steering_2023, privato_stacco_2024, liu_two_2025}. This way of treating the decoder as a generative model has been referred to as \textit{latent space navigation} or \textit{latent space walk} by practitioners in neighbouring fields \cite{akten_deep_2018}, and widely adopted in NIME design \cite{scurto_soundwalking_2023, symons_perceptron_2024}. However, as ``black box'' AI models, latent spaces do not encode musical controls in easy-to-understand ways \cite{bryan-kinns_exploring_2021, nabi_adding_2025}. Consequently, existing practices typically rely on the musician to ``blindly'' explore in the space and discover ways of using it \cite{nabi_embodied_2024}. In addition, general-purpose autoencoders can have a latent dimensionality that exceeds the upper limit of usable dimensions in an instrument \cite{levitin_control_2002}. These pose the challenge of designing interfaces with low-dimensional controls to access latent spaces, which we aim to address in this paper.

\subsection{Corpus-Based Sound Spaces}
\label{sec_soundspace}

Research in eXplainable AI (XAI) has explored ways of projecting the high-dimensional latent space into a low-dimensional control space \cite{bryan-kinns_exploring_2021} for accountable and steerable generation. A common approach is using \textit{Dimensionality Reduction} (DR) to extract low-dimensional representations of latent spaces. The use of DR can be seen in a range of works on corpus-based sound spaces \cite{schwarz_sound_2012, roma_adaptive_2019}, aiming to use sound descriptors to build and access a sound corpus and preserve essential characteristics of it \cite{jeon_unveiling_2025}. Recent works have explored the creation and interaction of sound spaces for a range of musical applications \cite{roma_adaptive_2019, garber_audiostellar_2020, tralie_concatenator_2024, ahn_eco-sonic_2025}. Other sound space mapping methods such as musical parameter mapping \cite{tubb_zoomable_2014} and latent mapping \cite{gregorio_augmenting_2019, murray-browne_latent_2021} are less relevant to our work because they do not emphasise the use of audio corpora.

The goal of our proposed sound space construction method, \anon[anonymous name]{Latent Terrain}, is related to but distinct from DR. It also aims to build low-dimensional control interfaces for high-dimensional latent spaces. However, rather than clustering latent codes in the interface as typical DR approaches do, we aim to construct ``maps'' for latent spaces that resemble the \textit{high-dynamic} 
nature of latent trajectories, which in effect offer higher spectral complexity for richer and more varied musical phrases.

\subsection{Material-Oriented DMI Design}

The applications of neural audio autoencoders in DMI design have focused on supporting functional tasks such as timbre transfer \cite{caspe_waveform_2025}, conditional synthesis \cite{demerle_combining_2024, nabi_adding_2025}, and text-to-audio \cite{evans_stable_2025}. These contrast our goal of exploring neural audio autoencoders as design materials, which we introduce below.

To position neural audio autoencoders as design materials, we take on Mudd's view of material-oriented musical interaction \cite{mudd_material-oriented_2019}, which treats materials’ resistance and constraints to be revealed and worked with to explore them as creative mediations \cite{gioti_composing_2022} for DMI design. This reflects Barad's notion of intra-action \cite[p.~189]{barad_meeting_2007}, which locates agencies in the entangled coupling between the designer and the technologies. Works that foreground this material-oriented view can embrace practice-related approaches that involve iterative crafting, prototyping, and experimenting with the material \cite{zappi_hackable_2018, shier_designing_2025}. For instance, early contextualisation of artistic practice during the technical development process \cite{armitage_tolvera_2024, shier_designing_2025}, or considering technical research as practice to be documented by first-person reflexive accounts \cite{pelinski_ways_2025}. 

Following this material-oriented view, we define our goal as treating autoencoders themselves as the site for exploring musical activities to understand their materiality in NIME.

\subsection{Material-Oriented Exploration of Neural Audio Autoencoders}

Various NIME examples illustrate how DMI designers unpack the materiality of neural audio autoencoders to explore their creative possibilities. For instance, the embodied and somatosensory aspects of latent representations have inspired works on movement-sound interactions \cite{nabi_embodied_2024, meyer_interactive_2025} and studies that look into performing techniques in latent spaces \cite{privato_stacco_2024, zheng_exploring_2025}, and the autonomous feature of latent spaces has been explored in live improvisations \cite{tahiroglu_ai-terity_2021, azimi_live_2025}, in which their uncertain and generative characteristics \cite{benjamin_machine_2021} are used as a source for music creativity. Moreover, various new interfaces for music composition \cite{tokui_latent_2025}, notation \cite{lepri_embodied_2024, bacon_approaches_2025, zheng_exploring_2025}, and sound design \cite{khallaghi_squishysonics_2025} have been explored. Our approach focuses on constructing sound spaces for latent spaces and their use in musical interface design.  

\section{Algorithm Design and Experiments}
\label{sec_algorithm}

Our exploration started by experimenting with the sound space construction method. This section focuses on technical challenges of the method, such as the mapping accuracy, underfitting and overfitting, and low-latency. We outline our design principles, algorithmic method, and data-driven experiments.

\subsection{Design Principles of \anon[Anonymous Name]{Latent Terrain}}
\label{sec_principles}

In Section~\ref{sec_soundspace} we reviewed notable approaches for constructing latent sound space, 
below we follow up with two interaction design principles that are unique in our work.

\subsubsection{Constructing sound spaces with latent trajectories}
\label{constructing}

\begin{figure}[t]
    \centering 
    \begin{subfigure}[b]{\linewidth}
      \centering
      \includegraphics[width=0.92\linewidth]{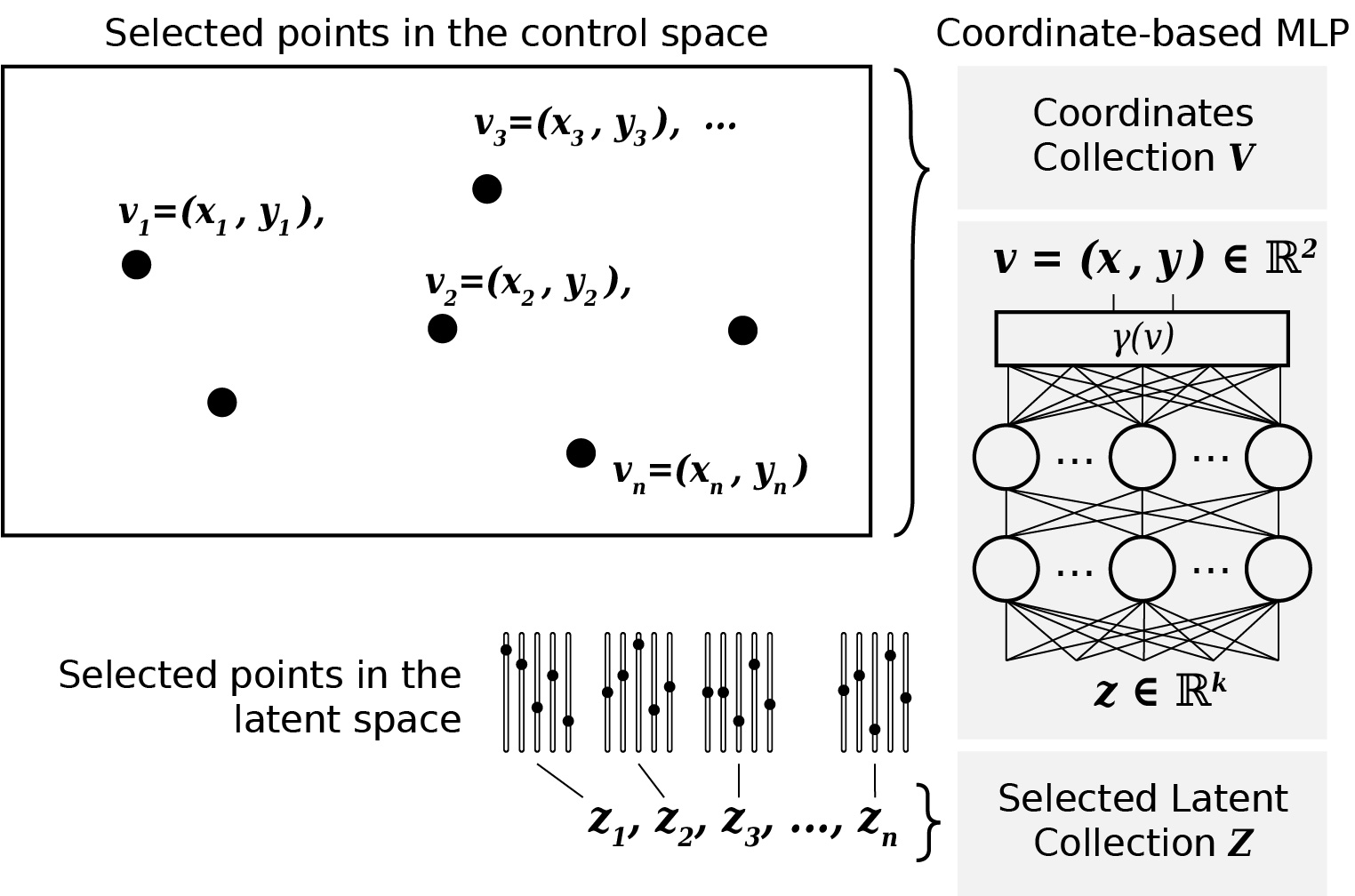}
      \caption{Vigliensoni and Fiebrink's method \citep{vigliensoni_steering_2023}.}
      \label{fig_diagram_vi}
    \end{subfigure}
    \begin{subfigure}[b]{\linewidth}
      \centering
      \includegraphics[width=0.92\linewidth]{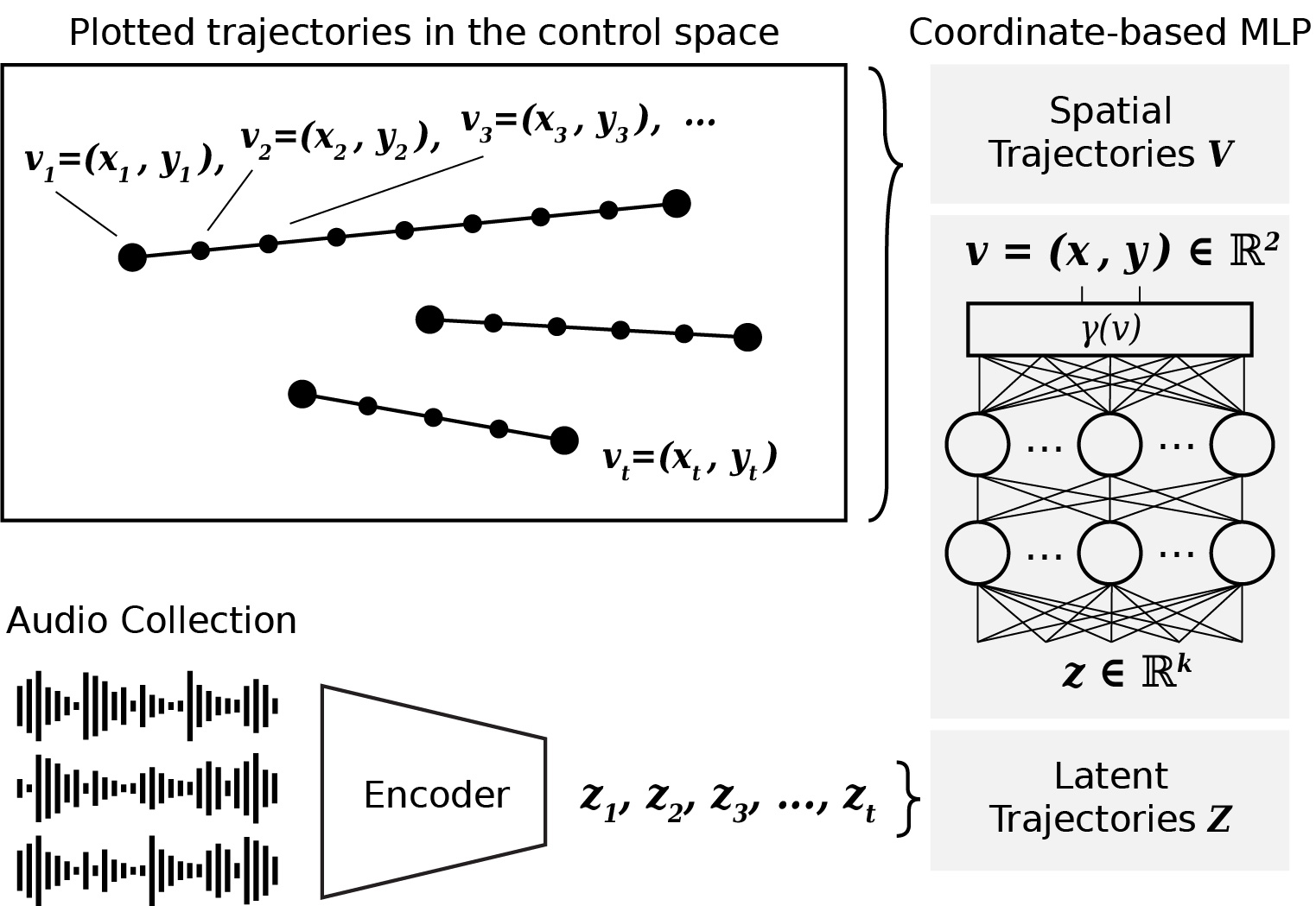}
      \caption{Our method supports the previous work \citep{vigliensoni_steering_2023}, and extends to modelling spatial trajectories paired with latent trajectories encoded from samples.}
      \Description{Diagram of the mapping model.}
      \label{fig_diagram}
    \end{subfigure}
\caption{Comparing the coordinate-to-latent mapping method by Vigliensoni and Fiebrink \citep{vigliensoni_steering_2023} with our method. 
}
\end{figure}

To offer intuitive controls over sound synthesis with latent spaces, we aim to create a coordinate-to-latent mapping that maps a control space to the latent space. Our initial design was inspired by the work of Vigliensoni and Fiebrink \cite{vigliensoni_steering_2023} (illustrated in Figure~\ref{fig_diagram_vi}). They used a Multi-Layer Perceptron (MLP) as the mapping model, which takes control space coordinates as inputs and is trained to output latent vectors. The core idea is that the MLP is trained with interactive machine learning \cite{fiebrink_meta-instrument_2009}.
In practice, the users explore and find interesting points in the latent space as targets, and select points in the control space that should map to these targets to create MLP training data \citep{vigliensoni_steering_2023}. At each timestep during inference, the trained MLP maps a control space coordinate to a latent vector, which is then reconstructed to audio waveforms by the decoder.

We extend the work of Vigliensoni and Fiebrink \cite{vigliensoni_steering_2023} from modelling latent vectors to modelling latent trajectories (see Section~\ref{sec_trajectories} for \textit{latent trajectory}). This enables the sound space to be constructed by long audio samples rather than single audio frames, therefore allowing longer musical phrases to be captured by the sound space. As shown in Figure~\ref{fig_diagram}, the training data of our MLP are (i) sequences of control space coordinates (\textbf{spatial trajectories}) plotted by the user, and (ii) sequences of latent vectors (\textbf{latent trajectories}) encoded from an audio collection, together forming the coordinate and latent pairs.

To state this problem precisely, Figure~\ref{fig_diagram} uses a 2-dimensional control space as an example. Given a neural audio autoencoder with $k$-dimensional latent space, a $c$-channel audio segment $x\in \mathbb{R}^{c\times T}$ can be encoded to a discrete latent sequence $Z = E(x) = \{z_1, z_2, ..., z_t\} \in \mathbb{R}^{k\times t}$, where $T$ and $t$ denote time at audio sampling rate and time at latent frame rate, respectively. The ratio between $T$ and $t$ is defined by the autoencoder's compression ratio. A discrete coordinate sequence $V = \{v_1, v_2, ..., v_t\} \in \mathbb{R}^{2\times t}$ in the control space is arranged by the user, where $v_t=(x_t, y_t) \in \mathbb{R}^2$. Our goal is to train a MLP model $f_\theta : \mathbb{R}^2 \to \mathbb{R}^k$ with paired $V$ and $Z$. Trained $f_\theta$ parameterises a latent trajectory as a function of continuous coordinates $v$. Models tackling similar tasks have been referred to as coordinate-based MLPs \cite{tancik_fourier_2020} or Compositional Pattern Producing Networks (CPPNs) \cite{stanley_compositional_2007, tesfaldet_fourier-cppns_2019}.

\begin{table*}[ht]
    \centering
    \small
    \caption{Our selection of autoencoders/codecs.}
    \begin{tabular}{lcccccccc}
        \toprule
        Autoencoder & Year & Latent dim. & Latent type$^5$ & Regularisation$^6$ & Comp.$^7$ & In Max$^8$ & Used in NIME design & \\
        \midrule
        RAVE$^1$ & 2021 & Various$^9$ & Continuous & Various$^9$ & 2048 & Yes & \scriptsize \cite{privato_magnetic_2023, scurto_soundwalking_2023, shepardson_living_2023, wilson_mosaick_2023, hein_transsonic_2024, nabi_embodied_2024, shaheed_i_2024, shepardson_tungnaa_2024, stefani_esteso_2024, symons_perceptron_2024, privato_stacco_2024, vigliensoni_data-_2024, visi_sophtar_2024, zheng_mapping_2024, bacon_approaches_2025, caspe_waveform_2025, khallaghi_squishysonics_2025, liu_two_2025, meyer_interactive_2025, mitcheltree_neutone_2025} \\
        Music2Latent$^2$ & 2024 & 64 & Continuous & Variational & 4096 & Yes & \scriptsize \cite{nistal_diff--riff_2024, tokui_latent_2025} \\
        Stable Audio Open$^3$ & 2024 & 64 & Continuous & Variational & 2048 & Yes & \scriptsize \cite{azimi_live_2025, nick_unstable_2025, swift_panic_2025} \\
        FlowDec$^4$ & 2025 & 128 & Discrete & RVQ \cite{kumar_high-fidelity_2023} & 1764 & No & \scriptsize None \\
        \bottomrule
        \multicolumn{9}{p{140mm}}{$^1$ Realtime Audio Variational autoEncoder (RAVE) \cite{caillon_rave_2021}, Max/MSP version available at \footnotesize\url{https://github.com/acids-ircam/RAVE}} \\
        \multicolumn{9}{p{165mm}}{$^2$ Music2Latent (M2L) \cite{pasini_music2latent_2024} is a consistency autoencoder. \anon[]{Our Max/MSP implementation: \footnotesize\url{https://github.com/jasper-zheng/music2latent-scripted}}} \\
        \multicolumn{9}{p{165mm}}{$^3$ The autoencoder in Stable Audio Open (SAO) 1.0 \cite{evans_stable_2025}. \anon[]{Our Max/MSP implementation: \footnotesize\url{https://github.com/jasper-zheng/streamable-stable-audio-open}}} \\
        \multicolumn{9}{p{165mm}}{$^4$ FlowDec \cite{welker_flowdec_2025} is a neural audio codec with discrete latent codes, with underlying codecs from \cite{kumar_high-fidelity_2023}. We only used the NDAC-25 configuration due to its low latent dimensionality and low latent frame rate (25 Hz).} \\
        \multicolumn{9}{p{165mm}}{$^5$ Latent type: Specifies whether the latent representations are in a continuous space or quantised into discrete codes.} \\
        \multicolumn{9}{p{165mm}}{$^6$ Regularisation: Defines how latent representations are learned during training, affects how information can be distributed in a latent space.} \\
        \multicolumn{9}{p{160mm}}{$^7$ Here we define Compression Ratio as the proportion by which the audio sampling rate is reduced to the latent frame rate, different from \cite{pasini_codicodec_2025}.} \\
        \multicolumn{9}{p{160mm}}{$^8$ Whether there is a version of that autoencoder implemented for realtime continuous inference in Max/MSP, by the time of writing.} \\
        \multicolumn{9}{p{165mm}}{$^9$ We used two RAVEs with Wasserstein regularisation and 16 latent dimensions, and one with variational regularisation and 4 latent dimensions.} \\
    \end{tabular}
    \label{tab_ae}
\end{table*}

\begin{figure}[h]
  \centering
  \includegraphics[width=0.97\linewidth]{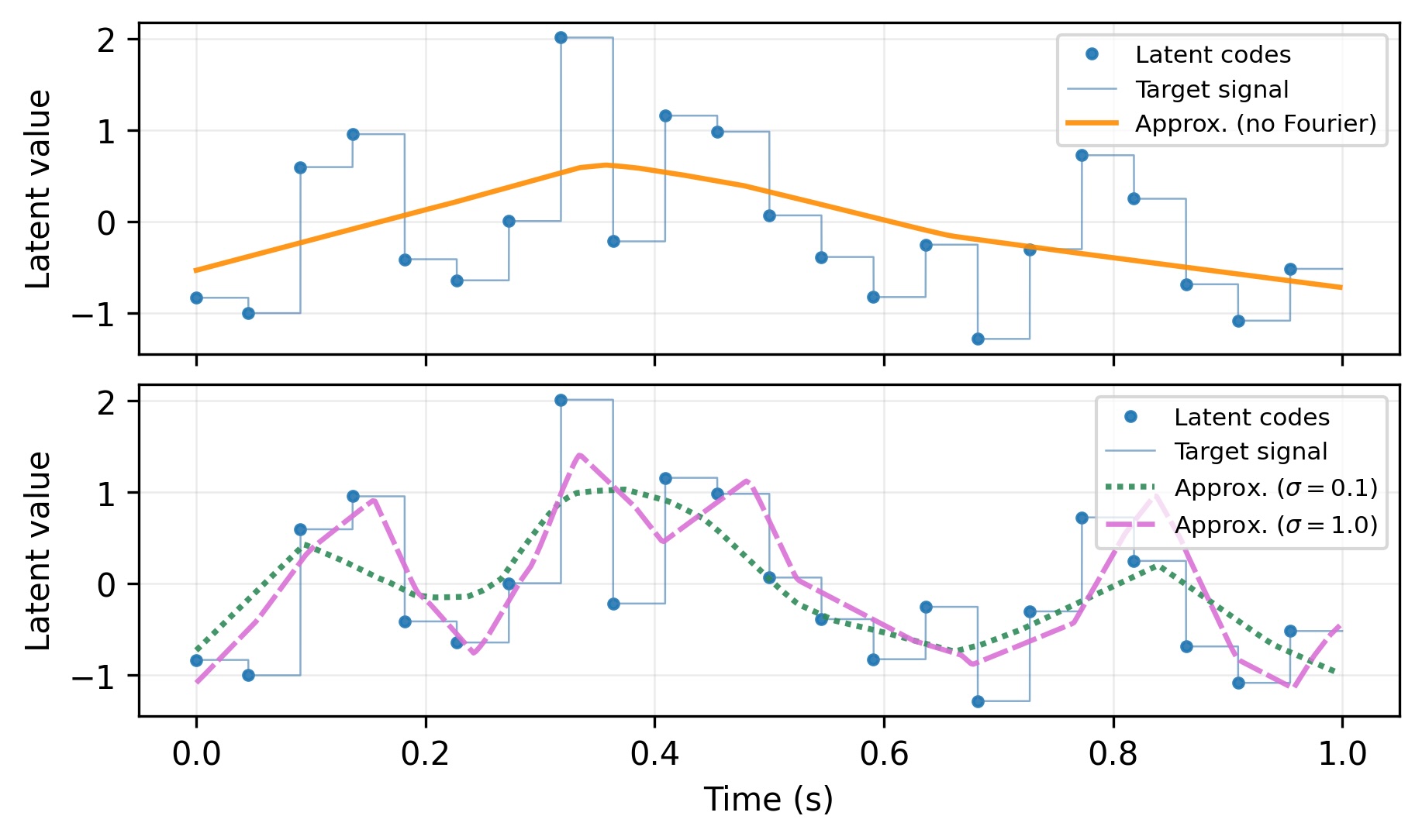}
  \caption{Latent codes of a 1-sec audio encoded by the autoencoder in Stable Audio Open \citep{evans_stable_2025}, showing the first latent dimension. Codes are converted to a latent signal by a zero-order hold. MLPs are trained to parameterised the signal given time $t$. A standard MLP (top: solid curve) have difficulty approximating the target signal. In contrast, MLPs with Fourier features (bottom: dashed and dotted curves) yield results that overcome ``spectral bias''.}
  \Description{The approximated latent trajectory with no Fourier failed to match the discrete latent signal. The approximated latent trajectory with $\sigma=1.0$ and $\sigma=1.0$ match the discrete latent signal. }
  \label{fig_approximation}
\end{figure}

\subsubsection{The high-frequency nature of latent trajectories}
\label{sec_frequency}

However, the practical challenge of training $f_\theta : \mathbb{R}^2 \to \mathbb{R}^k$ lies in the high-frequency nature of how audio samples traverse the latent space. The latent trajectory shown in Figure~\ref{fig_approximation} (top, dots) demonstrates that a sequence of vectors in a latent dimension can contain high-frequency components, in which latent values change rapidly. These components are crucial in synthesising phrases that have dynamically changing structure or timbre \citep{demerle_combining_2024}. However, when mapping these components to coordinates in dense Euclidean spaces, such as 1D linear timelines or 2D planes, standard MLPs can fail to model frequent changes in the sequence \cite{tancik_fourier_2020}, illustrated in Figure~\ref{fig_approximation} (top, the orange solid curve). This difficulty of approximating high-frequency components has been referred to as the phenomenon of \textbf{spectral bias} in machine learning literature \cite{rahaman_spectral_2019}. 
As a result, modelling latent trajectories in this way typically results in sound spaces that are overly smooth and lack musical dynamics. Although this way of smoothly interpolating between sounds has indeed fostered remarkable NIME designs \citep{symons_perceptron_2024, vigliensoni_data-_2024}, we hope to explore an alternative approach that can model high-dynamic musical phrases, to offer it as a variation to NIME designers.

\subsection{Modelling High-Frequency Latent Trajectories with Fourier Features}

Here we address the challenge formulated in Section~\ref{sec_principles}. 

\subsubsection{Fourier feature mapping}

Fourier feature mapping is a technique that allows neural networks to learn high-frequency functions in dense Euclidean coordinates \cite{tancik_fourier_2020}, proposed by Rahimi and Recht \cite{rahimi_random_2007}, primarily applied it in computer vision for image and geometry synthesis \cite{tesfaldet_fourier-cppns_2019}. Here we apply it to our coordinate-based MLP. 
Precisely, in a $d$-dimensional control space with coordinates $v\in \mathbb{R}^d$, we sample a matrix $B \in \mathbb{R}^{m\times d}$ from $\mathcal{N}(0, \sigma^2)$ with mapping size $m$ and Gaussian scale $\sigma$. Before passing $v$ to the MLP, we apply the Fourier feature mapping $\gamma(v)$:

\begin{equation}
\gamma (v) = \big[ \sin(2\pi B v),\; \cos(2\pi B v)\big]
\end{equation}

Equivalently, writing $b_k \in \mathbb{R}^d$ for the $k$-th row of $B$ shows that inputs in the control space is transformed by a series of sinusoidal functions with varying frequencies:

\begin{equation}
\begin{split}
\gamma(v) = \big[ \sin(2\pi b_1^\intercal v),\; \dots ,\; \sin(2\pi b_m^\intercal v), \\
\cos(2\pi b_1^\intercal v),\; \dots ,\; \cos(2\pi b_m^\intercal v) \big]
\end{split}
\end{equation}

In the following sections, we show our experiments on harnessing it for latent trajectory regression.

\subsubsection{Experiments and results}
\label{sec_exp}

We selected a range of autoencoders to form a plurality of materials, presented in Table~\ref{tab_ae}, aimed at covering those that (i) have open-source pre-trained models (ii) have been used in NIME for artistic practices, (iii) vary in configurations including dimensionality, regularisation, and compression ratio. 
We use the three autoencoders that have a Max/MSP version later in our NIME making.

To stay focused on the practical use of our mapping approach, we put the majority of our technical evaluation in the Appendix. 
In summary, shown in Table~\ref{tab_metrics}, Fourier features improve the mapping accuracy across all autoencoders, and the addition of the mapping MLP only minimally impacts the fidelity of reconstructed audio. Full details on the task, datasets, implementations,\footnote{The source code of all experiments is released at: \anon[anonymised for review]{\url{https://github.com/jasper-zheng/latent-terrain-pytorch}}} results, and discussion can be found in Appendix~\ref{apx_exp} and~\ref{apx_results}. We also inspect the training error with respect to multiple frequency components to verify that the Fourier features are enabling the MLP to learn high-frequency details as claimed, results can be found in Appendix~\ref{apx_freq}.

\begin{table}[t]
    \centering
    \caption{Regressing trajectories with the drum collection.}
    \label{tab_metrics}
    \begin{subtable}{\linewidth}
        \centering
        \footnotesize
        \caption{Peak Signal-to-Noise Ratio between original latent trajectories and regressed trajectories (higher is better).}
        \label{tab_psnr}
        \begin{tabular}{lcccc}
            \toprule
            & \multicolumn{4}{c}{\textbf{Peak Signal-to-Noise Ratio $\uparrow$}} \\
            \cmidrule(lr){2-5}
            Regression Config & RAVE & M2L & SAO & FlowDec \\
            \midrule
            No Fourier features & 27.516 & 29.774 & 21.697 & 23.530 \\
            Fourier features & \textbf{41.369} & \textbf{40.368} & \textbf{32.069} & \textbf{32.566} \\
            \bottomrule
        \end{tabular}
    \end{subtable}
    \\[0.75em]
    \begin{subtable}{\linewidth}
        \centering
        \footnotesize
        \caption{Fréchet Audio Distance of audio sets reconstructed from the original trajectories (encoder-decoder recon) and the regressed trajectories, comparing to the original audio set (lower is better).}
        \label{tab_fad}
        \begin{tabular}{lcccc}
            \toprule
            & \multicolumn{4}{c}{\textbf{Fréchet Audio Distance $\downarrow$}} \\
            \cmidrule(lr){2-5}
            Regression Config & RAVE & M2L & SAO & FlowDec \\
            \midrule
            Encoder-decoder recon. & 7.664 & 6.659 & 2.324 & 2.009 \\
            \midrule
            No Fourier features & 14.188 & 12.354 & 28.163 & 28.069 \\
            Fourier features & \textbf{8.139} & \textbf{7.511} & \textbf{2.387} & \textbf{2.473} \\
            \bottomrule
        \end{tabular}
    \end{subtable}
\end{table}

\begin{figure*}[t]
  \centering
  \includegraphics[width=\linewidth]{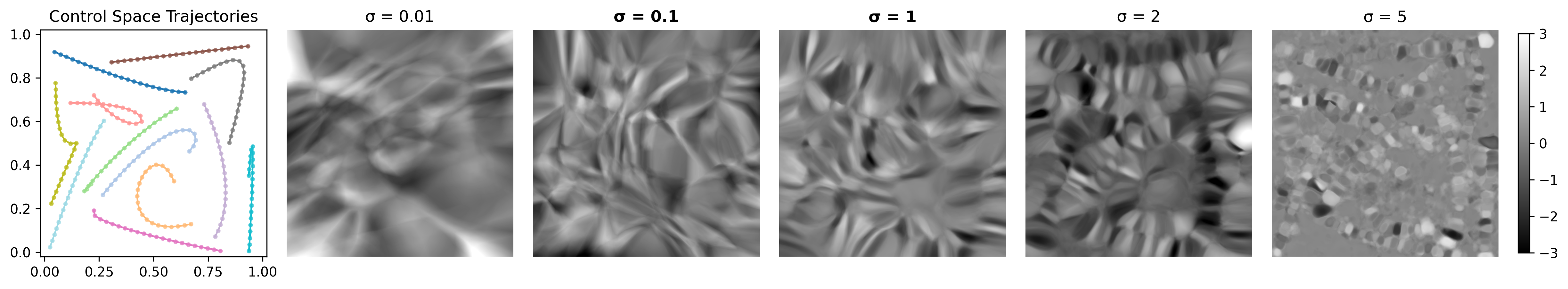}
  \caption{\anon[Anonymous name]{Latent terrains} created by pairing controls space trajectories (randomly sampled) and latent trajectories (encoded by SAO using a set of 1-sec drum samples from MUSDB18 \cite{rafii_musdb18-hq_2019}). Visualisations of magnitude at the first latent dimension with various Gaussian scales $\sigma$. Lower $\sigma$ causes overly flattened interpolation (underfit). Larger $\sigma$ causes noisy sound space and poor generalisation to areas not covered by training data (overfit). This is subject to experimentation, but we found $\sigma \in [0.1, 1]$ generally can lead to a musically interesting result.}
  \Description{Control space trajectories and sound space visualisation created from different sigma value}
  \label{fig_sigma}
\end{figure*}


\subsection{Configuring Fourier Features in Practice}
\label{sec_tuning}

Here we experiment with the mapping model with a 2D control plane with coordinates $(x, y) \in [0, 1]^2$, which is a common setting in the practice of sound space \cite{schwarz_sound_2012}. As shown in Figure~\ref{fig_sigma}, we train the mapping MLP with random control space trajectories (in practice these should be defined interactively by the user) paired with latent trajectories. We then visualise the sound space by sampling the MLP with coordinates in the $[0, 1]^2$ interval and display the first latent dimension as a greyscale matrix. 

The Gaussian scale $\sigma$ is a key parameter in configuring Fourier features when constructing the sound space. This mirrors the findings in \cite{tancik_fourier_2020} that $\sigma$ qualitatively affects the balance between underfit and overfit. Visulisations in Figure~\ref{fig_sigma} illustrate this point by showing the effect of $\sigma$ on the overall flatness of the sound space. As described in the caption, a proper $\sigma$ leads to \anon[an anonymous]{a terrain} with a steep and mountainous surface, which in effect leads to spectrally complex and varied sound when navigating the control space. We therefore expose this parameter to the API (see Section~\ref{sec_situating}) to defer the practical parameter tuning to makers and designers who will be creating the sound space.

\subsection{On-the-fly Training and Inference}
\label{sec_fly}

Relevant works on neural audio typically evaluate model efficiency at inference \citep{caspe_waveform_2025}. However, in our setting the interactive training also plays an important role in adapting autoencoders. We therefore also evaluate the on-device training speed. 

\begin{figure}[t]
  \centering
  \includegraphics[width=\linewidth]{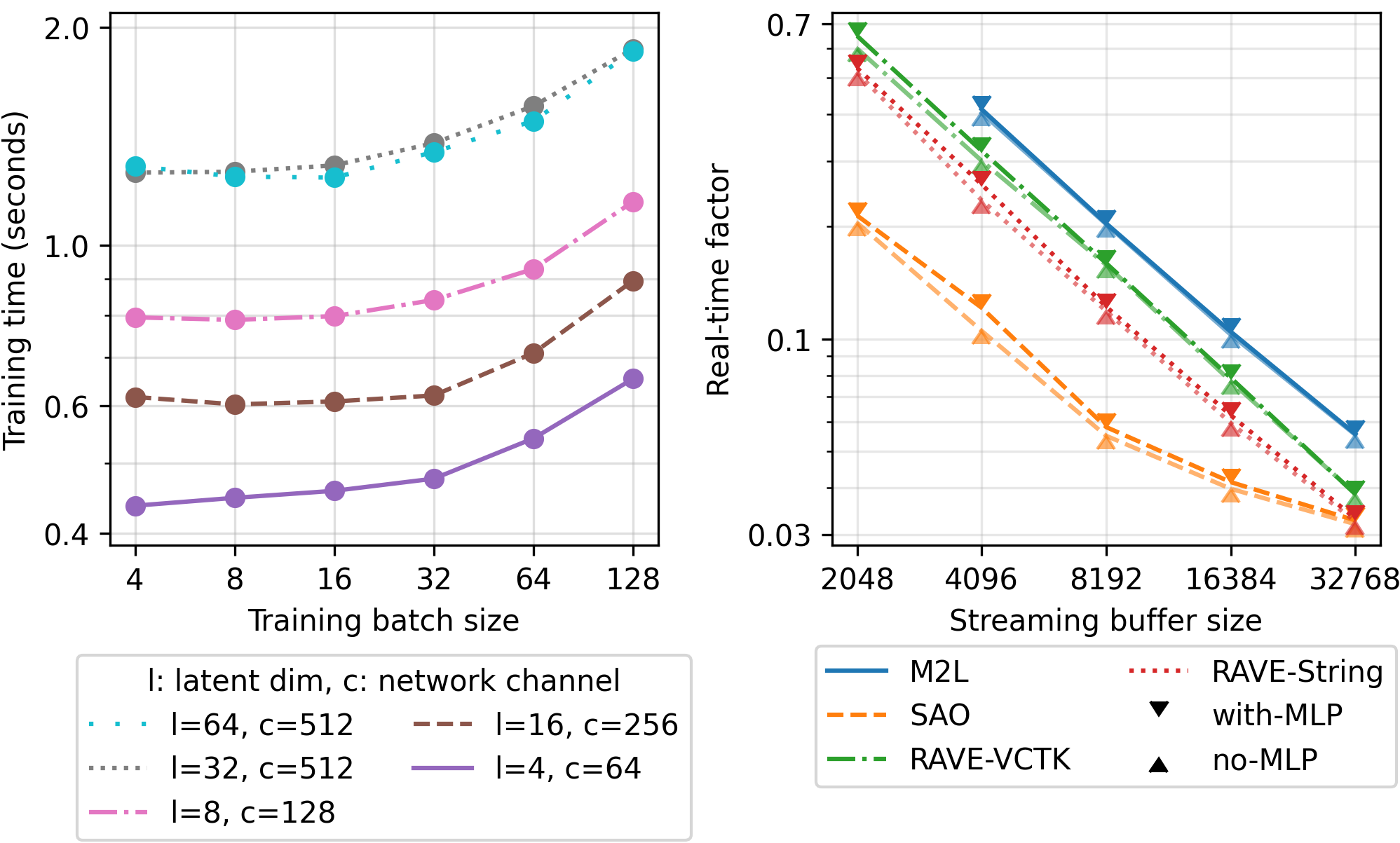}
  \caption{Left: Average time needed to train the MLP for 1000 steps against training batch size, across MLP configurations. Right: Mean real-time factor against buffer size, across all decoders, comparing with and without the MLP. Mean values were computed over 100 runs per condition. Both metrics are lower the better.}
  \label{fig_time}
\end{figure}

We measured the time taken to train the mapping MLP for 1000 steps across various training batch sizes. A training step is defined as one optimisation cycle in which one batch of training data is processed. 
Results in Figure~\ref{fig_time} (Left) were tested on the CPU of an Apple Silicon M4 Max laptop with 48GB RAM. Training in all conditions can be done within 2 seconds. To put this into context, encoding a 10-sec audio sample results in 215 latent codes. With a batch size of 16, one training epoch can be done by 14 steps. In our experience, 500 epochs can ensure a converged training, therefore yields roughly 1.2 minutes training time. This matches works such as \citep{shier_real-time_2024} that involve training MLPs as mapping models, claiming an under 2 minutes training time.

For real-time sound synthesis, we measured the four autoencoders across various streaming buffer sizes by Real-Time Factor (RTF), calculated as the ratio between the time taken to process a buffer and the temporal duration of that buffer. We compared RTFs under two conditions: inference of the decoder together with the mapping MLP, and inference of only the decoder with randomly sampled latent vectors. Results in Figure~\ref{fig_time} (Right) were tested on the GPU of the same M4 Max laptop, showing that the addition of the MLP only introduces a minimum increase in the RTF. Full results across more devices are shown in Appendix~\ref{apx_fly}.

\section{Situating \anon[Project Name]{Latent Terrain} in DMI Design}
\label{sec_situating}

We developed \texttt{\anon[project.name]{nn.terrain$\sim$}},\footnote{\anon[Anonymous link, an anonymised copy of some webpages can be found in Supplementary Material.]{\url{https://jasper-zheng.github.io/nn_terrain/}}} a package of Max/MSP objects for NIME makers, encapsulating the sound space method described in Section~\ref{sec_algorithm}. We refer to the sound space as \textit{\anon[anonymous]{terrain}} to draw an analogy to \anon[anonymous name synth]{wave terrain synthesis \cite{james_developing_2005}}. Objects in the package provide Max-style APIs that allow users to build, adapt, and use \anon[anonymous]{terrains} with pre-trained autoencoders. The package also includes utility objects for interactive trajectory plotting, real-time latent recording, trajectory playback, and more.

The \texttt{\anon[project.name]{nn.terrain$\sim$}} online project page provides installation guide, API documentation, instructions, tutorials, and Max help files. It also includes artist-contributed patch examples and demos. We hope to use this collectively built repository to respond to the conference theme \textbf{Communities}. In particular, the musical and technical communities that are exploring AI-enhanced DMIs such as neural audio autoencoders \cite{caillon_rave_2021, mitcheltree_neutone_2025, noel-hirst_sampling_2025}. 


\subsection{Origin Story}

To situate the algorithm into the environment of DMI design, the development of \texttt{\anon[project.name]{nn.terrain$\sim$}} is informed by the first author's practices of music compositions with neural audio autoencoders. The API implementation is shaped by the author's intended ways of building sound spaces and using them for music compositions in practice. At the time of writing, the author had been using \texttt{\anon[project.name]{nn.terrain$\sim$}} for DMI design for 12 months, from the initial prototype to the publicly released package. This continuous technical practice \cite{pelinski_ways_2025} is documented by write-ups and presentations at a series of public engagement events across the start, middle, and end of the project, combined with major technical change logs (shown in Appendix~\ref{apx_milestones}). Here we report three design milestones that focus on the minutiae and rationale of appropriating the technical artefact into creative practices. The aim is to demonstrate how musical interactions can emerge from, and be shaped by, the material-oriented resistances and constraints.

\subsection{Milestones and Design Artefacts}
\label{sec_milstones}

We describe three Graphical User Interface (GUI) designed along with the package development. These interfaces mark the milestones that lead to the latest release of \texttt{\anon[project.name]{nn.terrain$\sim$}}. All GUIs (screenshots in Figure~\ref{fig_interfaces}) explore ways of using \texttt{\anon[project.name]{nn.terrain$\sim$}} for music compositions and performances. 

\subsubsection*{Interface A: Embodied Sound Space Navigation}

The first GUI design with \texttt{\anon[project.name]{nn.terrain$\sim$}} focuses on exploring the embodied movement-sound coupling in the sound space. It offers a drawing pad as the control space, in which the stylus $(x,y)$ location is used to navigate in the sound space and produce real-time sonic feedback. The artistic intent behind the interface is to use the drawing canvas as a space to create graphical music scores, and to trace the musical gestures in the scores. 
One of the emerging strategies of using it is annotating sonically interesting zones on the drawing pad and experimenting with ways of re-enacting sounds in these zones. 

\subsubsection*{Interface B: Interactive Sound Space Construction}

In the second design and development phase, we explore the use of \anon[anonymous]{terrains} in a wider range of models and audio collections. Therefore the API design focuses on supporting the interactive \anon[anonymous]{terrain} construction within the Max/MSP environment. As a result, the cycle of training pairs gathering, model training, monitoring, and inference was distributed to three interdependent Max objects. 
One emerging way of using the \anon[anonymous]{terrain} for sound synthesis is looking for ways of programming the \anon[anonymous]{terrain} navigation. For instance, we added the right-most panel that allows modulating the stylus $(x,y)$, mimicking existing practices of composing with sound spaces, such as the ``oscillations and orbits'' control in \cite{roma_agent-based_2023}.

\subsubsection*{Interface C: Programming Trajectory Playback}

\begin{figure}[t]
  \centering
  \includegraphics[width=\linewidth]{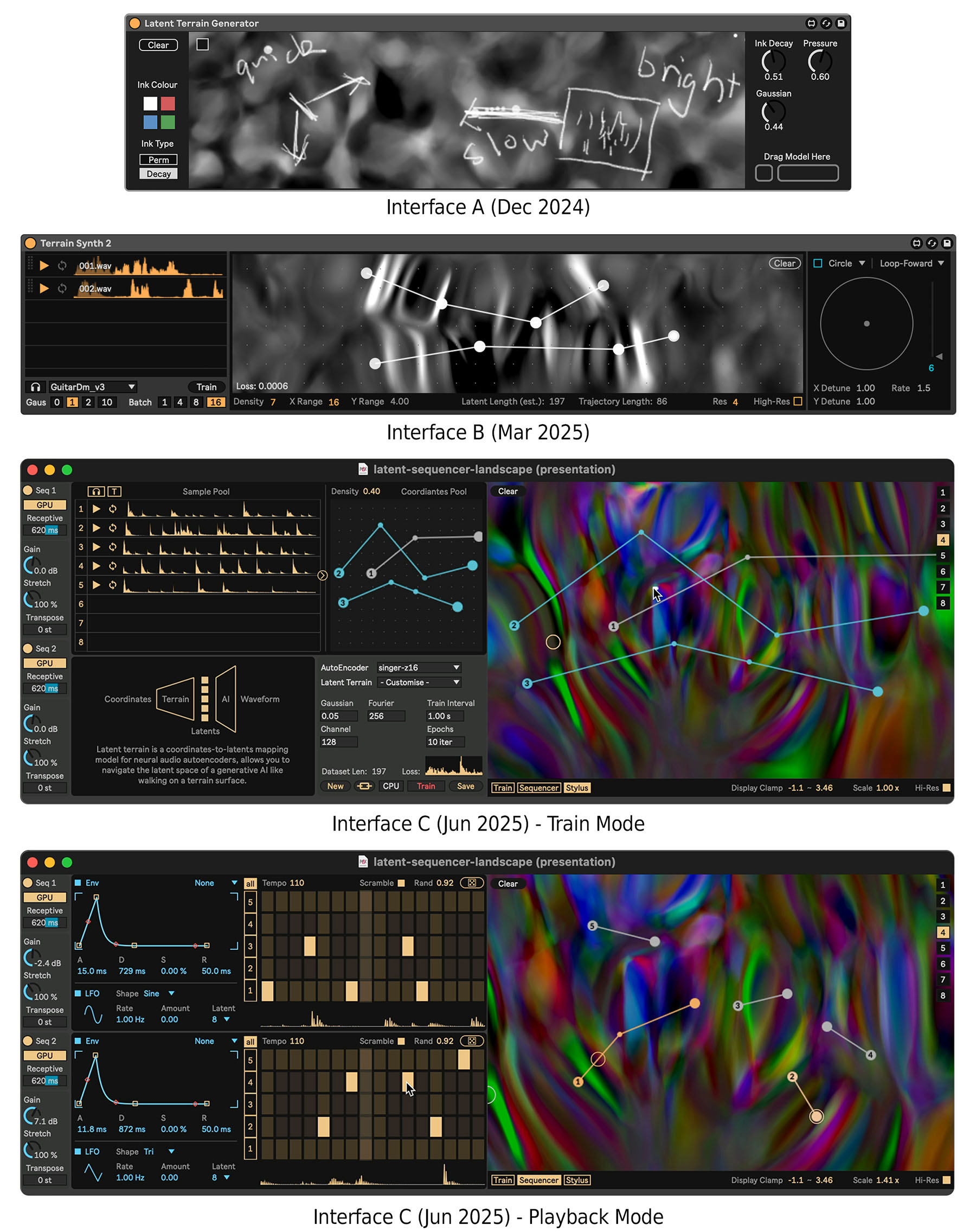}
  \caption{Three GUI designs with \anon[Anonymous Name]{Latent Terrain}, span from Dec 2024 to Jun 2025. Recorded demos of each interface can be found in the online project page.}
  \Description{Three GUI designs for music composition.}
  \label{fig_interfaces}
\end{figure}

In the third phase, we explore more ways of navigating the \anon[anonymous]{terrains} in sample-based music composition. The input control space coordinates were exposed to the API to allow for customised input devices, rather than constrained to the stylus drawing pad to open the design space of input devices. \anon[An \texttt{anonymous.gui}]{An \texttt{nn.terrain.gui}} object is refined as a utility object to control the playback of trajectories. As shown in Figure~\ref{fig_interfaces}C, the user first loads audio samples and plots trajectories in the ``train mode'' the same way as the previous interface. Then in the ``playback mode'' they define a new set of trajectories from the 2D canvas as ``latent samples''. Trajectories can be programmed by the pattern in the step sequencer on the left. 

\section{Practice-Based Exploration}
\label{sec_artistic}


After the initial release of the \texttt{\anon[project.name]{nn.terrain$\sim$}} package, we collaborated with four artist-researchers to explore NIME design with autoencoder and \anon[project name]{latent terrain}. The connection with each artist-researcher is motivated by a mutual interest in exploring neural audio synthesis in NIME. We aimed to situate \texttt{\anon[project.name]{nn.terrain$\sim$}} within the practice of each artist-researcher \citep{sturdee_plurality_2021}, in particular, audio programming, music performance, DMI, and sound installation design.\footnote{Detailed positionality statement of each artist can be found in Appendix~\ref{apx_artists}.} The collaborations started in the summer of 2025 and lasted over three to five months. During the period, we had regular meetings (either bi-weekly or monthly, with a combination of remote and in-person modes) with each artist-researcher to catch up on the progress. The ideation and conceptualisation of each work are done by the artist-researcher after their first meeting with the first author to grasp an initial understanding of the package. Whereas the role of the first author is a facilitator who briefs and explains the use of \texttt{\anon[project.name]{nn.terrain$\sim$}}, signposts relevant resources and documentation, and provides technical support throughout the process. The artist-researchers join this paper as the second to fifth co-authors to bring their voices into our discussion and acknowledge their project contribution.

\begin{table*}[t]
    \centering
    \footnotesize
    \caption{Summary of the four projects in our portfolio.}
    \label{tab_psnr}
    \begin{tabular}{p{0.9cm}p{3.0cm}p{1.9cm}p{4.1cm}p{5.5cm}}
        \toprule
         & \textbf{Type of work} & \textbf{Autoencoder} & \textbf{Audio collection} & \textbf{Key aspect explored by the work} \\
        \midrule
        Project 1 & A data sonification patch           & RAVE                    & Samples of vintage music & 
        Interactive sound space adaptation by biofeedback  \\
        
        Project 2 & An audio-visual environment & Stable Audio Open       & Archival and field recordings in Hutongs & 
        Crafting a sound space for an virtual gallery  \\
        
        Project 3 & A live performance patch        & RAVE                & Samples of organ instrument recordings & 
        Autonomous micro-variations in latent space navigation \\
        
        Project 4 & A soundscape composition & Stable Audio Open   & Various (drum loops, vocals, and piano) & 
        Neural synthesis as a resampling tool for composition    \\
        \bottomrule
    \end{tabular}
    \label{tab_projects}
\end{table*}

\subsection{Portfolio of Works}

To collectively showcase artist-researchers’ ways of integrating \texttt{\anon[project.name]{nn.terrain$\sim$}} into NIME design, we create an annotated portfolio. An annotated portfolio \citep{gaver_annotated_2012} brings together design artefacts into a systematic body of work, to share the design thinking and experience behind them, and show the broader design issues the collection addresses. 
The following sections and Table~\ref{tab_projects} present a summary of each project. The full portfolio is also showcased on the project website.\footnote{\anon[Anonymous link, an anonymised copy of some webpages can be found in Supplementary Material.]{\url{https://jasper-zheng.github.io/nn_terrain/posts/annotated-portfolio/}}}

\subsubsection*{\textbf{\anon[Artist and Project Name 1]{Project 1: Repressive Terrain} by \anon[ArtistName1]{Keigo Yoshida}}} 
\label{sec_p1}

\begin{figure}[h]
  \centering
  \includegraphics[width=\linewidth]{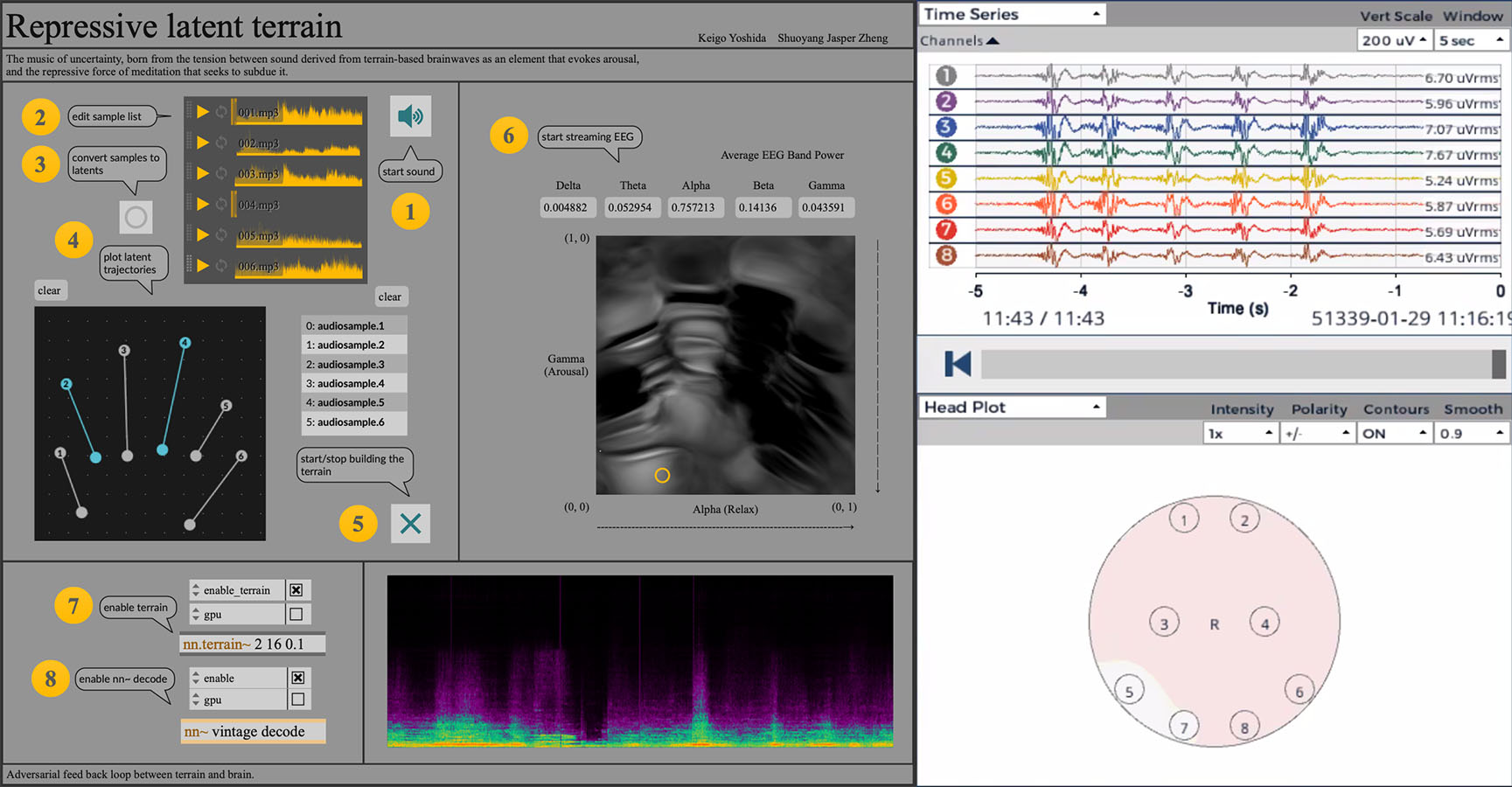}
  \caption{\anon[Project Name 1 by Artist 1]{Screenshots of the customised patch created by Keigo Yoshida and the first author, using with signals from an OpenBCI EEG headset}.}
  \Description{Exploring the latent space through an OpenBCI Electroencephalography (EEG) headset.}
  \label{fig_keigo}
\end{figure}

A data sonification patch (Figure~\ref{fig_keigo}) that aims to create a meditative listening experience, in which the audience's time-varying Electroencephalography (EEG) data is sonified to a real-time synthesised soundscape. The artistic theme is the active manipulation of consciousness in a focused meditation, and an adversarial tension between the audience's striving for calm and the algorithm's impulse toward arousal sonic responses. It exploits the interactive sound space adaptation aspect of the \anon[Anonymised Name]{Latent Terrain} toolkit to create a real-time evolving soundscape.

\subsubsection*{\textbf{\anon[Artist and Project Name 2]{Project 2: nn/mémoire} by \anon[ArtistName2]{Jiatong Liu}}} 
\label{sec_p2}
\begin{figure}[t]
  \centering
  \includegraphics[width=\linewidth]{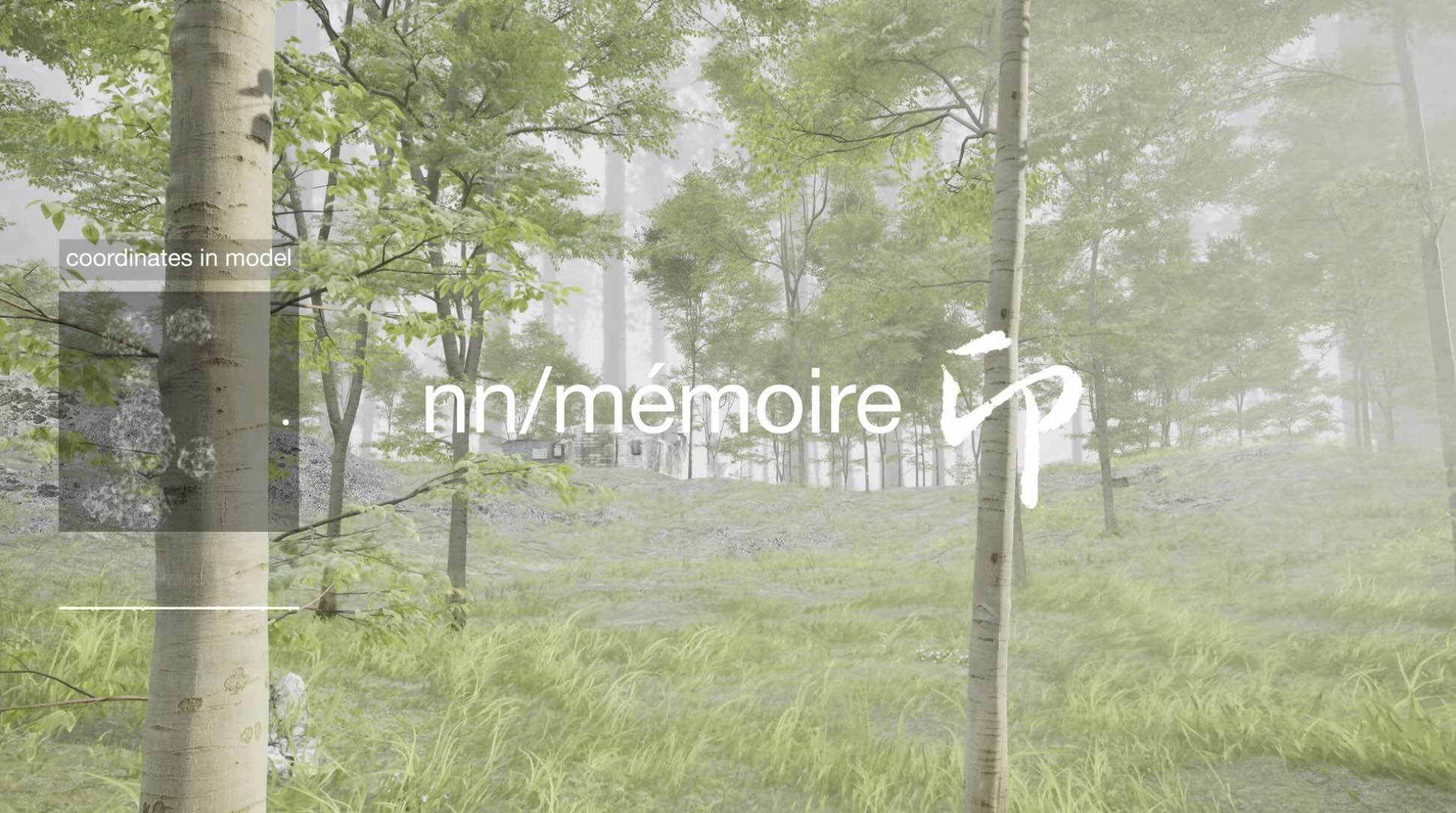}
  \caption{\anon[Project Name 2 by Artist 2]{A screenshot of nn/mémoire by Jiatong Liu}.}
  \Description{An audio-visual virtual environment.}
  \label{fig_jt}
\end{figure}

An audio-visual virtual environment (Figure~\ref{fig_jt}) that aims to create an embodied listening experience that reimagines the sound of cultural heritage in Hutongs, a type of traditional northern Chinese courtyard house in Beijing. The environment is a virtual gallery that allows audiences to wander through and hear the real-time generated soundscape that captures the sound of Hutongs. The sound space is constructed with archival recordings from the Sound Art Museum Beijing, aiming to preserve the vanishing living culture that is now vanishing due to urban changes. It explores methods to craft latent spaces into an immersive sound environment and how latent spaces open a new approach to sound design.

\subsubsection*{\textbf{\anon[Project Name 3 by Artist 3 and Collaborator]{Project 3: Trek by Nico García-Peguinho and Nikhil Bullock}}} 
\label{sec_p3}
A collaborative composition and live performance system that aims for controlled timbral materials creation from minimal gestural input. It explores sampling and interpolation strategies for the latent space. Specifically, coordinates in the latent space are indexed into zones based on their acoustic characteristics. It then uses a boids algorithm\footnote{\url{https://cs.stanford.edu/people/eroberts/courses/soco/projects/2008-09/modeling-natural-systems/boids.html}} to create micro-variations to navigate these zones, to generate naturalistic sonic drift. It uses post-synthesis signal processing techniques to sculpt the sound. The autoencoder is in duet with a generative drum patch.

\begin{figure}[t]
  \centering
  \includegraphics[width=\linewidth]{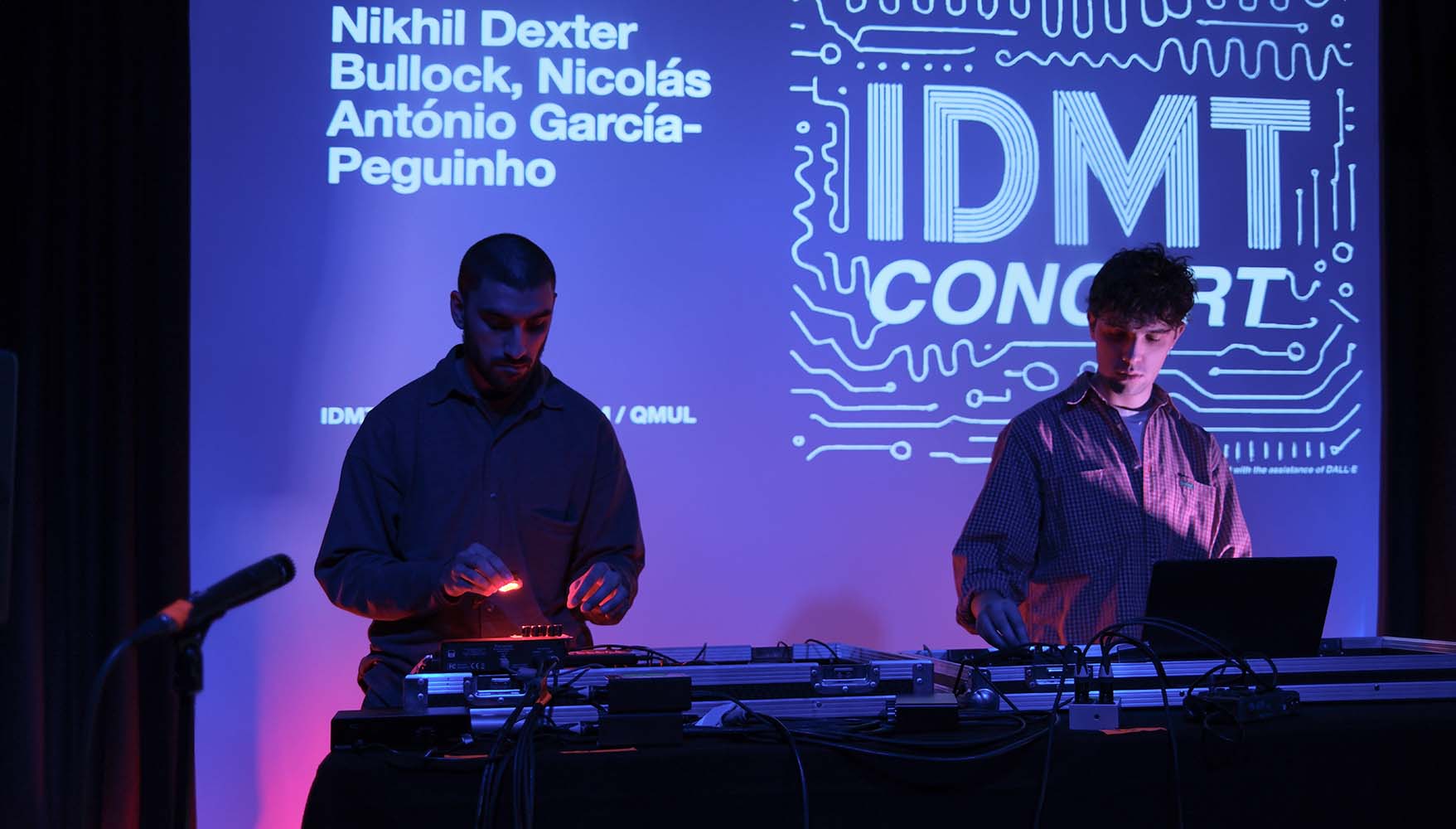}
  \caption{\anon[Project Name 3 by Artist 3 and Collaborator]{Nico García-Peguinho and Nikhil Bullock} performing their piece at the \anon[One Anonymised Concert Name]{Interactive Digital Multimedia Techniques Concert} at \anon[One Anonymised University Name]{Queen Mary University of London}.
  }
  \Description{.}
  \label{fig_nico}
\end{figure}

\begin{figure}[h]
  \centering
  \includegraphics[width=\linewidth]{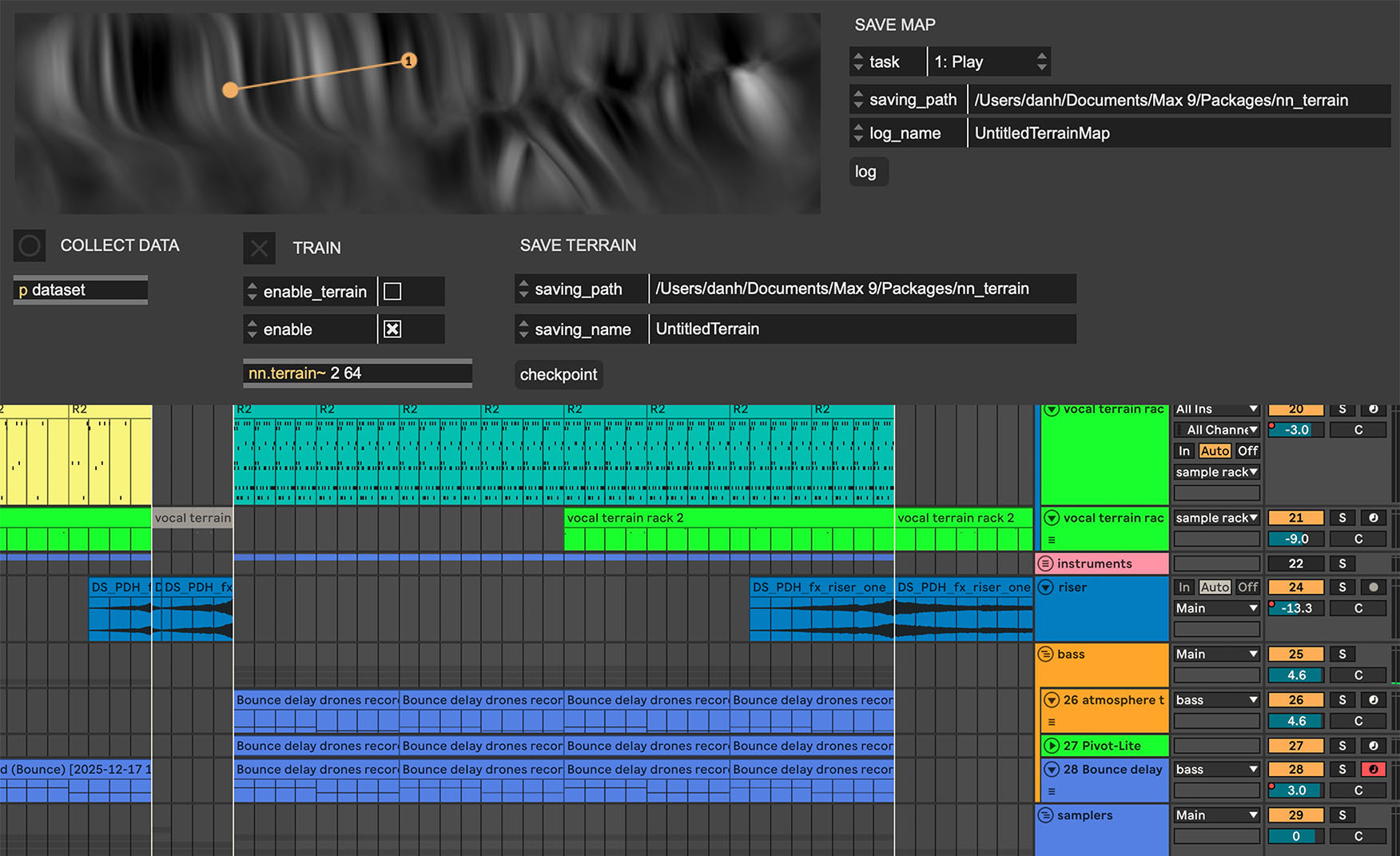}
  \caption{Screenshots showing the cutomised Max/MSP patch by \anon[Artist 4 Name]{Dan Hearn}.
  }
  \Description{Screenshots showing the cutomised Max/MSP patch, AI audio materials are used with non-AI music composition tools.}
  \label{fig_dan}
\end{figure}

\subsubsection*{\textbf{\anon[Artist and Project Name 4]{Project 4: ambient\_terrain\_1 by Dan Hearn}}} 
\label{sec_p4}
A soundscape composition produced by resampling recorded audio and field recordings. \anon[Artist 4]{Dan} explores ways of adapting autoencoders to his own existing music production practice. He created a customised patch to generate short audio loops by sampling in the latent space. Audio materials derived from the patch are used throughout the track with non-AI composition tools. This composition process explores the combination of AI and non-AI materials in ways that the use of an AI tool is not immediately apparent to listeners, but seamlessly integrated in the sample-based music production workflow.

\subsection{Annotated Themes}
\label{sec_annotations}

To elicit the design considerations and artistic choices raised by each project, we conducted a semi-structured interview with each artist-researcher after the completion of their project. The interviews are either 60-min live or asynchronous, focusing on artist-researchers describing the concept and matter of concern of their work
. We then transcribed the interview audio recordings to analyse emerging themes that bridge between their work and broader research opportunities in NIME, Human-Computer Interaction (HCI), and AI, which could inform future works. Here we present these themes.

\subsubsection*{\textbf{Uncertainty and ambiguity as characteristics}} 

Across NIME and HCI, uncertainty and ambiguity have been considered as constitutive qualities that can be engaged by artistic practice \citep{caramiaux_explorers_2022, bryan-kinns_using_2024}, requiring artists to deliberately work and cultivate their practices to respond to the unpredictability \citep{tahiroglu_-terity_2020}. In the making process of all four pieces, the uncertainty and ambiguity features of neural audio autoencoders have been discussed repeatedly. ``Learning to deal with the unpredictability'' (\anon[Artist2]{Jiatong}) has become a central aspect to be considered in the interface design. For instance, \anon[Artist4]{Dan} characterised his workflow as ``explorative and serendipitous'', \anon[Artist3]{Nico} referred to an ``active listening'' \citep{noel-hirst_sampling_2025} way of composing in which ambiguity remained but became legible through embodied exploring and listening. 
In \anon[Artist1]{Keigo}'s piece, the constantly changing sound space is seen as an adversarial entity that actively resisted the performer's intentions, turning uncertainty into a compositional driver. In this respect, the complex and opaque aspects of latent spaces is acknowledged, and captured as a source for uncertainty and ambiguity within design processes.

\subsubsection*{\textbf{Entangled agencies in material-centred design}}
In a material-centred view, the computational materiality of AI is considered as an active medium \citep{scurto_prototyping_2021}, in which creative practice is shaped and constrained by materials rather than solely directed by human intention imposed upon tools \citep{wiberg_material-centered_2018}. In musical interaction, such perspectives 
locate agencies in the engagement and negotiation between the designer and their materials \citep{mudd_material-oriented_2019}. In our portfolio, 
\anon[Artist2]{Jiatong} describes that ``[the design of the virtual gallery space] reveals itself'' through the movement and listening in the latent space, whereas \anon[Artist1]{Keigo} describes ``let it traces and accesses past states (training data) in real-time, not to establish full control over the outcome''. Taken together, design intent arise not from transparent control over the materials, but from learning how the material's capacities act, respond, and transform the creative process.

\subsubsection*{\textbf{Designing with a material assemblage}} 
Across all four pieces in our portfolio, the autoencoder itself is not an isolated instrument. Instead, makers worked with a material constellation in which the autoencoders interact with other materials in the workspace, forming an ecology of processes and environment \citep{rodger_what_2020}. Such as \anon[Artist2]{Jiatong}’s virtual gallery space, which was built according to the latent sound space. In this respect, design intents arise not from the autoencoder alone, but from its entanglement with tools, bodies, and environments. From an \textit{assemblage} view \citep{gioti_composing_2022} of AI, autoencoders belong to a broader constellation of human and non-human actors, including other synthesis algorithms, sensors, other coding environments, and the listening and composing practices. This assemblage is not a static configuration of materials, but an ongoing process in which materials are continuously constructed and de(con)structed \citep{bowers_hybrid_2014}. For instance, \anon[Artist4]{Dan} described that ``[autoencoder] gradually finds its place within my sample-based music production workflow''--the role of autoencoders in the creative practice became clear through repeated trials and errors, rather than a predetermined identity. 



\section{Discussion}
\label{sec_discussion}

The material-oriented exploration presented in this paper is not about inventing new neural audio autoencoders, nor proposing new model steering strategies that surpass works such as \citep{vigliensoni_steering_2023, demerle_combining_2024, nabi_adding_2025}. Instead, it aims to use the \anon[Anonymous Name]{Latent Terrain} package to position neural audio autoencoders within a space of NIME crafting, where they are engaged as materials to be probed and learned through practice. Here we discuss our response to the research question: \textit{What are the challenges and opportunities of using neural audio autoencoders as design materials for NIME makers?} 

\subsection{Latent Spaces as Tailorable Sound Spaces}

We highlight two aspects of neural audio autoencoders that were explored by the makers to discuss their opportunities in NIME making.

First, makers explore ways of constructing the latent sound space that align with their creative intent. \anon[Anonymous Name]{Latent Terrain} can be seen as a way of adapting autoencoders with a curated audio collection. In the broader space of AI as material for artworks, building and curating training datasets has been an approach for artists to intervene in generative models and bring agency into their work \citep{caramiaux_explorers_2022, arzberger_reflexive_2024}. \anon[Anonymous Name]{Latent Terrain} offers an alternative way for artists to integrate their curated audio collection: tailoring the latent spaces of large generative models into a smaller and customised subset, in the form of a sound space. 
The use of curated datasets was explored in \anon[Artist2]{Jiatong}'s work, in which the choice of using archival recordings as the audio corpus aligns with the themes of the vanishing living culture. In addition, \anon[Artist1]{Keigo}'s work explored the interactive adaptation of the sound space, in which the mapping model is constantly being trained and updated, forming the theme of uncertainty in the listening experience.

Second, makers explore ways of navigating the latent sound space to balance control and uncertainty. \anon[Anonymous Name]{Latent Terrain} can be seen as an approach of tailoring the high-dimensional latent space into a low-dimensional control space. In our portfolio, ways of navigating the latent sound space are entry points to make clear the connection between the autoencoder and a range of non-AI materials in the makers' existing workspace or practices. For instance, \anon[Artist 4]{Dan}'s practice of sample-based music composition informed his way of sampling in the latent space to derive materials for composition, whereas \anon[Artist 3]{Nico}'s use of the boids algorithm to navigate the space has become a unique theme in his work. 

In summary, we suggest that future work explore the use of curated audio collections, the interactive adaptation of latent space, and the plural ways of navigating the latent sound space to discover its opportunities in NIME.

\subsection{Toward a Shared View on Autoencoders}
\label{sec_implications}

In Section~\ref{sec_intro} we highlighted several practical challenges of enabling hands-on engagement with neural audio autoencoders such as technical barriers, lack of resources and support, and the complex nature of latent spaces. Reflecting on the development of \texttt{\anon[project.name]{nn.terrain$\sim$}} and the artistic exploration, we argue that the lack of a space to share practice, knowledge, and techniques can be another challenge in unpacking the creative possibilities of autoencoders in NIME making. 

When collecting the four pieces into the portfolio, each maker uses domain-specific techniques to work with autoencoders, such as the ``active and embodied listening'' approach by \anon[Artist 2]{Jiatong} and \anon[Artist 3]{Nico}. They also invent unique vocabularies to articulate their technical practices, such as \anon[Artist 4]{Dan}'s way of describing his latent resampling approach and \anon[Artist 1]{Keigo}'s way of describing the synthesis process as ``tracing and accessing past state''. We suggest that these multifaceted approaches can benefit from a sustained community that establishes shareable practices \citep{Wenger_1998}, technical know-how \citep{tremblay_enabling_2021}, and folk theories \citep{shelby_generative_2024} to explore AI materials in NIME making. 
In a broader DMI design view, cultural exchanges with and within artists' communities play an important role in supporting digital luthiers' practice-based engagement with technical artefacts \citep{armitage_studying_2023}. An example can be drawn from how the turntable was transformed into a musical instrument by the development of its own musical culture, techniques, and virtuosi \citep[p.~326]{jorda_instruments_2004}. Neural audio autoencoders, on a similar trajectory, become design materials in NIME through sustained, practice-based engagement within a crafting space that is shared by both music communities and technical communities.

\section{Conclusion}

We contribute a technical method, a Max/MSP package, and a practice-based account that foregrounds crafting as a site to explore emerging AI-enhanced NIME designs. We focused on the decoder-only latent space walk technique with neural audio autoencoders. Our technical experimentation with autoencoders resulted in \anon[Anonymous Name]{Latent Terrain}, a Max/MSP package that constructs low-dimensional sound spaces from high-dimensional latent trajectories. By positioning neural audio autoencoders within a crafting space, we explored the challenges and opportunities of using them as design materials in NIME. Through a collaborative exploration with four artist-researchers, we delivered an annotated portfolio to showcase a domain of design opened by neural audio autoencoders. We highlighted that makers explore ways of constructing the latent sound space that align with their creative intent, and ways of navigating them to establish control over the sound synthesis. We called for a community view of autoencoders in which artists and makers are empowered to share technical know-how and practices with AI materials to unpack their creative possibilities in NIME making.


\section{Ethical Standards}
All autoencoder models used in the papers are pre-trained models trained on audio files licensed under CC0, CC BY, or CC Sampling Plus. Data used for interactive machine learning is either under CC0 or CC BY licences, or created by the authors themselves. 

The practice-based exploration described in the paper is conducted within the authors and did not involve experiments with other human participants. The project was approved by \anon[the name of an anonymous ethics committee]{the Queen Mary University of London ethics committee}, reference number: \texttt{QMERC20.565.DSEECS25.026}.


\begin{acks}
\anon{Shuoyang Zheng} is a research student supported by \anon{the EPSRC UKRI Centre for Doctoral Training in Artificial Intelligence and Music (grant number EP/S022694/1)}. We would like to thank all the artists for their time and engagement. Thank you to Joseph Meyer and all colleagues who have gave thoughtful feedback to help improving the quality of this paper.

\end{acks}

\balance

\bibliographystyle{ACM-Reference-Format}
\bibliography{reference-corrected.bib}

@inproceedings{roma_adaptive_2019,
	address = {Porto Alegre, Brazil},
	title = {Adaptive mapping of sound collections for data-driven musical interfaces},
	url = {http://www.nime.org/proceedings/2019/nime2019_paper060.pdf},
	doi = {10.5281/zenodo.3672976},
	booktitle = {Proceedings of the {International} {Conference} on {New} {Interfaces} for {Musical} {Expression}},
	publisher = {UFRGS},
	author = {Roma, Gerard and Green, Owen and Tremblay, Pierre Alexandre},
	editor = {Xambó Sedó, Anna},
	month = jun,
	year = {2019},
	pages = {313--318},
}

@inproceedings{rodger_what_2020,
	address = {Birmingham, UK},
	title = {What makes a good musical instrument? {A} matter of processes, ecologies and specificities},
	url = {https://www.nime.org/proceedings/2020/nime2020_paper79.pdf},
	doi = {10.5281/zenodo.4813438},
	booktitle = {Proceedings of the {International} {Conference} on {New} {Interfaces} for {Musical} {Expression}},
	publisher = {Birmingham City University},
	author = {Rodger, Matthew and Stapleton, Paul and van Walstijn, Maarten and Ortiz, Miguel and Pardue, Laurel S},
	editor = {Michon, Romain and Schroeder, Franziska},
	month = jul,
	year = {2020},
	pages = {405--410},
}

@article{jorda_instruments_2004,
	title = {Instruments and players: {Some} thoughts on digital lutherie},
	volume = {33},
	issn = {0929-8215},
	shorttitle = {Instruments and {Players}},
	url = {https://doi.org/10.1080/0929821042000317886},
	doi = {10.1080/0929821042000317886},
	number = {3},
	urldate = {2024-01-02},
	journal = {Journal of New Music Research},
	author = {Jordà, Sergi},
	month = sep,
	year = {2004},
	pages = {321--341},
}

@inproceedings{fiebrink_meta-instrument_2009,
	address = {Pittsburgh, PA, United States},
	title = {A meta-instrument for interactive, on-the-fly machine learning},
	url = {http://www.nime.org/proceedings/2009/nime2009_280.pdf},
	doi = {10.5281/zenodo.1177513},
	booktitle = {Proceedings of the {International} {Conference} on {New} {Interfaces} for {Musical} {Expression}},
	author = {Fiebrink, Rebecca and Trueman, Dan and Cook, Perry R.},
	year = {2009},
	pages = {280--285},
}

@misc{intelligent_instruments_lab_rave-models_2023,
	title = {rave-models ({Revision} ad15daf)},
	url = {https://huggingface.co/Intelligent-Instruments-Lab/rave-models},
	publisher = {Hugging Face},
	author = {{Intelligent Instruments Lab}},
	year = {2023},
	doi = {10.57967/hf/1235},
}

@article{armitage_studying_2023,
	title = {Studying {Subtle} and {Detailed} {Digital} {Lutherie}: {Motivational} {Contexts} and {Technical} {Needs}},
	issn = {2220-4806},
	journal = {Proceedings of the International Conference on New Interfaces for Musical Expression},
	author = {Armitage, Jack and Magnusson, Thor and McPherson, Andrew},
	editor = {Ortiz, Miguel and Marquez-Borbon, Adnan},
	month = may,
	year = {2023},
	pages = {1--9},
}

@inproceedings{scurto_soundwalking_2023,
	address = {Mexico City, Mexico},
	title = {Soundwalking {Deep} {Latent} {Spaces}},
	doi = {10.5281/zenodo.11189166},
	booktitle = {Proceedings of the {International} {Conference} on {New} {Interfaces} for {Musical} {Expression}},
	author = {Scurto, Hugo and Postel, Ludmila},
	editor = {Ortiz, Miguel and Marquez-Borbon, Adnan},
	month = may,
	year = {2023},
	pages = {232--235},
}

@inproceedings{shepardson_tungnaa_2024,
	address = {Utrecht, Netherlands},
	title = {Tungnaá: a {Hyper}-realistic {Voice} {Synthesis} {Instrument} for {Real}-{Time} {Exploration} of {Extended} {Vocal} {Expressions}},
	issn = {2220-4806},
	url = {https://doi.org/10.5281/zenodo.13904943},
	doi = {10.5281/zenodo.13904943},
	booktitle = {Proceedings of the {International} {Conference} on {New} {Interfaces} for {Musical} {Expression}},
	author = {Shepardson, Victor and Reus, Jonathan and Magnusson, Thor},
	editor = {Bin, S. M. Astrid and Reed, Courtney N.},
	month = sep,
	year = {2024},
	pages = {536--540},
}

@book{Wenger_1998, series={Learning in Doing: Social, Cognitive and Computational Perspectives}, title={Communities of Practice: Learning, Meaning, and Identity}, publisher={Cambridge University Press}, author={Wenger, Etienne}, year={1998}, collection={Learning in Doing: Social, Cognitive and Computational Perspectives} }

@inproceedings{pasini_codicodec_2025,
	title = {{CoDiCodec}: {Unifying} {Continuous} and {Discrete} {Compressed} {Representations} of {Audio}},
	url = {https://doi.org/10.5281/zenodo.17811403},
	doi = {10.5281/zenodo.17811403},
	booktitle = {Proceedings of the 26th {International} {Society} for {Music} {Information} {Retrieval} {Conference}},
	publisher = {ISMIR},
	author = {Pasini, Marco and Lattner, Stefan and Fazekas, George},
	month = sep,
	year = {2025},
	pages = {447--455},
}

@inproceedings{tahiroglu_latent_2024,
	address = {Sweden},
	title = {Latent {Spaces} as {Platforms} for {Sonic} {Creativity}},
	urldate = {2024-11-01},
	booktitle = {Proceedings of the 15th {International} {Conference} on {Computational} {Creativity}},
	author = {Tahiroğlu, Koray and Wyse, Lonce},
	year = {2024},
}

@incollection{yee-king_latent_2022,
	address = {Cham},
	title = {Latent {Spaces}: {A} {Creative} {Approach}},
	isbn = {978-3-031-10960-7},
	url = {https://doi.org/10.1007/978-3-031-10960-7_8},
	booktitle = {The {Language} of {Creative} {AI}: {Practices}, {Aesthetics} and {Structures}},
	publisher = {Springer International Publishing},
	author = {Yee-King, Matthew},
	editor = {Vear, Craig and Poltronieri, Fabrizio},
	year = {2022},
	pages = {137--154},
}

@article{tubb_zoomable_2014,
	title = {A {Zoomable} {Mapping} of a {Musical} {Parameter} {Space} {Using} {Hilbert} {Curves}},
	volume = {38},
	issn = {01489267, 15315169},
	doi = {10.1162/COMJ_a_00254},
	number = {3},
	urldate = {2024-10-28},
	journal = {Computer Music Journal},
	author = {Tubb, Robert and Dixon, Simon},
	year = {2014},
	pages = {23--33},
}

@article{privato_stacco_2024,
	title = {Stacco: {Exploring} the {Embodied} {Perception} of {Latent} {Representations} in {Neural} {Synthesis}},
	issn = {2220-4806},
	url = {http://nime.org/proceedings/2024/nime2024_62.pdf},
	doi = {10.5281/zenodo.13904899},
	journal = {Proceedings of the International Conference on New Interfaces for Musical Expression},
	author = {Privato, Nicola and Shepardson, Victor and Lepri, Giacomo and Magnusson, Thor},
	editor = {Bin, S. M. Astrid and Reed, Courtney N.},
	month = sep,
	year = {2024},
	pages = {424--431},
}

@inproceedings{sturdee_plurality_2021,
	address = {New York, NY, USA},
	series = {C\&{C} '21},
	title = {A {Plurality} of {Practices}: {Artistic} {Narratives} in {HCI} {Research}},
	isbn = {978-1-4503-8376-9},
	url = {https://doi.org/10.1145/3450741.3466771},
	doi = {10.1145/3450741.3466771},
	booktitle = {Proceedings of the 13th {Conference} on {Creativity} and {Cognition}},
	publisher = {Association for Computing Machinery},
	author = {Sturdee, Miriam and Lewis, Makayla and Strohmayer, Angelika and Spiel, Katta and Koulidou, Nantia and Alaoui, Sarah Fdili and Urban Davis, Josh},
	year = {2021},
}

@inproceedings{scurto_prototyping_2021,
	address = {New York, NY, USA},
	series = {{DIS} '21},
	title = {Prototyping {Machine} {Learning} {Through} {Diffractive} {Art} {Practice}},
	isbn = {978-1-4503-8476-6},
	url = {https://doi.org/10.1145/3461778.3462163},
	doi = {10.1145/3461778.3462163},
	booktitle = {Proceedings of the 2021 {ACM} {Designing} {Interactive} {Systems} {Conference}},
	publisher = {Association for Computing Machinery},
	author = {Scurto, Hugo and Caramiaux, Baptiste and Bevilacqua, Frederic},
	year = {2021},
	pages = {2013--2025},
}

@inproceedings{wilson_mosaick_2023,
	address = {Utrecht, Netherlands},
	title = {{MosAIck}: {Staging} {Contemporary} {AI} {Performance} - {Connecting} {Live} {Coding}, {E}-{Textiles} and {Movement}},
	shorttitle = {{MosAIck}},
	url = {https://zenodo.org/records/7843540},
	doi = {10.5281/zenodo.7843540},
	urldate = {2024-07-29},
	booktitle = {Proceedings of the 7th {International} {Conference} on {Live} {Coding}},
	publisher = {Zenodo},
	author = {Wilson, Elizabeth and Schubert, Deva and Satomi, Mika and McLean, Alex and Amaya Gozalez, Juan Felipe},
	month = apr,
	year = {2023},
	doi = {10.5281/zenodo.7843540},
}

@inproceedings{vigliensoni_steering_2023,
	address = {Ontario, Canada},
	title = {Steering latent audio models through interactive machine learning},
	doi = {10.5281/zenodo.8087978},
	booktitle = {In {Proceedings} of the 14th {International} {Conference} on {Computational} {Creativity}},
	author = {Vigliensoni, Gabriel and Fiebrink, Rebecca},
	year = {2023},
}

@inproceedings{tahiroglu_-terity_2020,
	address = {Birmingham, UK},
	title = {Al-terity: {Non}-{Rigid} {Musical} {Instrument} with {Artificial} {Intelligence} {Applied} to {Real}-{Time} {Audio} {Synthesis}},
	url = {https://www.nime.org/proceedings/2020/nime2020_paper65.pdf},
	doi = {10.5281/zenodo.4813402},
	booktitle = {Proceedings of the {International} {Conference} on {New} {Interfaces} for {Musical} {Expression}},
	publisher = {Birmingham City University},
	author = {Tahiroğlu, Koray and Kastemaa, Miranda and Koli, Oskar},
	editor = {Michon, Romain and Schroeder, Franziska},
	month = jul,
	year = {2020},
	pages = {337--342},
}

@inproceedings{zheng_mapping_2024,
	title = {A {Mapping} {Strategy} for {Interacting} with {Latent} {Audio} {Synthesis} {Using} {Artistic} {Materials}},
	doi = {10.48550/arXiv.2407.04379},
	booktitle = {Proceedings of {The} second international workshop on {eXplainable} {AI} for the {Arts} ({XAIxArts} 2)},
	author = {Zheng, Shuoyang and Xambó Sedó, Anna and Bryan-Kinns, Nick},
	year = {2024},
}

@inproceedings{tesfaldet_fourier-cppns_2019,
	title = {Fourier-{CPPNs} for {Image} {Synthesis}},
	url = {https://ieeexplore.ieee.org/document/9022247/?arnumber=9022247},
	doi = {10.1109/ICCVW.2019.00392},
	urldate = {2024-07-15},
	booktitle = {2019 {IEEE}/{CVF} {International} {Conference} on {Computer} {Vision} {Workshop} ({ICCVW})},
	author = {Tesfaldet, Mattie and Snelgrove, Xavier and Vazquez, David},
	month = oct,
	year = {2019},
	pages = {3173--3176},
}

@article{roma_agent-based_2023,
	title = {Agent-{Based} {Music} {Live} {Coding}: {Sonic} adventures in {2D}},
	volume = {28},
	doi = {10.1017/S1355771823000274},
	number = {2},
	journal = {Organised Sound},
	author = {Roma, Gerard},
	year = {2023},
	pages = {231--240},
}

@inproceedings{murray-browne_latent_2021,
	title = {Latent {Mappings}: {Generating} {Open}-{Ended} {Expressive} {Mappings} {Using} {Variational} {Autoencoders}},
	shorttitle = {Latent {Mappings}},
	url = {https://nime.pubpub.org/pub/latent-mappings/release/1},
	doi = {10.21428/92fbeb44.9d4bcd4b},
	urldate = {2023-11-26},
	booktitle = {International {Conference} on {New} {Interfaces} for {Musical} {Expression}},
	author = {Murray-Browne, Tim and Tigas, Panagiotis},
	month = jun,
	year = {2021},
}

@inproceedings{bryan-kinns_exploring_2021,
	title = {Exploring {XAI} for the {Arts}: {Explaining} {Latent} {Space} in {Generative} {Music}},
	booktitle = {1st {Workshop} on {eXplainable} {AI} approaches for debugging and diagnosis},
	author = {Bryan-Kinns, Nick and Banar, Berker and Ford, Corey and Reed, Courtney N and Zhang, Yixiao and Colton, Simon and Armitage, Jack},
	year = {2021},
}

@inproceedings{bryan-kinns_using_2024,
	address = {New York, NY, USA},
	series = {C\&{C} '24},
	title = {Using {Incongruous} {Genres} to {Explore} {Music} {Making} with {AI} {Generated} {Content}},
	isbn = {979-8-4007-0485-7},
	doi = {10.1145/3635636.3656198},
	booktitle = {Proceedings of the 16th {Conference} on {Creativity} \& {Cognition}},
	publisher = {Association for Computing Machinery},
	author = {Bryan-Kinns, Nick and Noel-Hirst, Ashley and Ford, Corey},
	year = {2024},
	pages = {229--240},
}

@article{zappi_hackable_2018,
	title = {Hackable {Instruments}: {Supporting} {Appropriation} and {Modification} in {Digital} {Musical} {Interaction}},
	volume = {5},
	issn = {2297-198X},
	shorttitle = {Hackable {Instruments}},
	url = {https://doi.org/10.3389/fict.2018.00026},
	doi = {10.3389/fict.2018.00026},
	urldate = {2025-01-04},
	journal = {Frontiers in ICT},
	author = {Zappi, Victor and McPherson, Andrew},
	month = oct,
	year = {2018},
}

@book{goodfellow_deep_2016,
	title = {Deep {Learning}},
	publisher = {MIT Press},
	author = {Goodfellow, Ian and Bengio, Yoshua and Courville, Aaron},
	year = {2016},
}

@article{shaheed_i_2024,
	title = {I {Am} {Sitting} in a ({Latent}) {Room}},
	issn = {2220-4806},
	url = {http://nime.org/proceedings/2024/nime2024_49.pdf},
	doi = {10.5281/zenodo.13904872},
	journal = {Proceedings of the International Conference on New Interfaces for Musical Expression},
	author = {Shaheed, Nicholas and Wang, Ge},
	editor = {Bin, S. M. Astrid and Reed, Courtney N.},
	month = sep,
	year = {2024},
	pages = {333--338},
}

@inproceedings{lepri_embodied_2024,
	address = {New York, NY, USA},
	series = {{AM} '24},
	title = {Embodied {Sketching} for {Neural} {Synthesis}},
	isbn = {979-8-4007-0968-5},
	url = {https://doi.org/10.1145/3678299.3678358},
	doi = {10.1145/3678299.3678358},
	booktitle = {Proceedings of the 19th {International} {Audio} {Mostly} {Conference}: {Explorations} in {Sonic} {Cultures}},
	publisher = {Association for Computing Machinery},
	author = {Lepri, Giacomo and Privato, Nicola and Magnusson, Thor},
	year = {2024},
	pages = {549--551},
}

@article{pelinski_ways_2025,
	title = {Ways of knowing, ways of writing: technical practice research in new musical instrument design},
	volume = {53},
	url = {https://doi.org/10.1080/09298215.2024.2442348},
	doi = {10.1080/09298215.2024.2442348},
	number = {0},
	journal = {Journal of New Music Research},
	author = {Pelinski, Teresa and McPherson, Andrew and Fiebrink, Rebecca},
	year = {2025},
	pages = {79--92},
}

@article{armitage_tolvera_2024,
	title = {Tölvera: {Composing} {With} {Basal} {Agencies}},
	issn = {2220-4806},
	url = {http://nime.org/proceedings/2024/nime2024_42.pdf},
	doi = {10.5281/zenodo.13904854},
	journal = {Proceedings of the International Conference on New Interfaces for Musical Expression},
	author = {Armitage, Jack and Shepardson, Victor and Magnusson, Thor},
	editor = {Bin, S. M. Astrid and Reed, Courtney N.},
	month = sep,
	year = {2024},
	pages = {282--291},
}

@inproceedings{rahaman_spectral_2019,
	address = {Long Beach, California},
	series = {Proceedings of {Machine} {Learning} {Research}},
	title = {On the {Spectral} {Bias} of {Neural} {Networks}},
	volume = {97},
	booktitle = {Proceedings of the 36th {International} {Conference} on {Machine} {Learning}},
	publisher = {PMLR},
	author = {Rahaman, Nasim and Baratin, Aristide and Arpit, Devansh and Draxler, Felix and Lin, Min and Hamprecht, Fred and Bengio, Yoshua and Courville, Aaron},
	editor = {Chaudhuri, Kamalika and Salakhutdinov, Ruslan},
	month = jun,
	year = {2019},
	pages = {5301--5310},
}

@inproceedings{nabi_embodied_2024,
	address = {New York, NY, USA},
	series = {{MOCO} '24},
	title = {Embodied exploration of deep latent spaces in interactive dance-music performance},
	isbn = {979-8-4007-0994-4},
	url = {https://doi.org/10.1145/3658852.3659072},
	doi = {10.1145/3658852.3659072},
	booktitle = {Proceedings of the 9th {International} {Conference} on {Movement} and {Computing}},
	publisher = {Association for Computing Machinery},
	author = {Nabi, Sarah and Esling, Philippe and Peeters, Geoffroy and Bevilacqua, Frédéric},
	year = {2024},
}

@inproceedings{jeon_unveiling_2025,
	address = {New York, NY, USA},
	title = {Unveiling {High}-dimensional {Backstage}: {A} {Survey} for {Reliable} {Visual} {Analytics} with {Dimensionality} {Reduction}},
	shorttitle = {Unveiling {High}-dimensional {Backstage}},
	url = {http://arxiv.org/abs/2501.10168},
	doi = {10.1145/3706598.3713551},
	urldate = {2025-04-16},
	booktitle = {Proceedings of the {SIGCHI} {Conference} on {Human} {Factors} in {Computing} {Systems} ({CHI} '25)},
	publisher = {Association for Computing Machinery},
	author = {Jeon, Hyeon and Lee, Hyunwook and Kuo, Yun-Hsin and Yang, Taehyun and Archambault, Daniel and Ko, Sungahn and Fujiwara, Takanori and Ma, Kwan-Liu and Seo, Jinwook},
	year = {2025},
}

@inproceedings{tancik_fourier_2020,
	address = {Red Hook, NY, USA},
	series = {{NIPS} '20},
	title = {Fourier features let networks learn high frequency functions in low dimensional domains},
	isbn = {978-1-7138-2954-6},
	booktitle = {Proceedings of the 34th {International} {Conference} on {Neural} {Information} {Processing} {Systems}},
	publisher = {Curran Associates Inc.},
	author = {Tancik, Matthew and Srinivasan, Pratul P. and Mildenhall, Ben and Fridovich-Keil, Sara and Raghavan, Nithin and Singhal, Utkarsh and Ramamoorthi, Ravi and Barron, Jonathan T. and Ng, Ren},
	year = {2020},
}

@inproceedings{bryan-kinns_xaixarts_2025,
	address = {New York, NY, USA},
	series = {{CHI} {EA} '25},
	title = {{XAIxArts} {Manifesto}: {Explainable} {AI} for the {Arts}},
	isbn = {979-8-4007-1395-8},
	url = {https://doi.org/10.1145/3706599.3716227},
	doi = {10.1145/3706599.3716227},
	booktitle = {Proceedings of the {Extended} {Abstracts} of the {CHI} {Conference} on {Human} {Factors} in {Computing} {Systems}},
	publisher = {Association for Computing Machinery},
	author = {Bryan-Kinns, Nick and Zheng, Shuoyang Jasper and Castro, Francisco and Lewis, Makayla and Chang, Jia-Rey and Vigliensoni, Gabriel and Broad, Terence and Clemens, Michael Paul and Wilson, Elizabeth},
	year = {2025},
}

@article{vigliensoni_data-_2024,
	title = {Data- and interaction-driven approaches for sustained musical practices with machine learning},
	volume = {53},
	issn = {0929-8215},
	url = {https://doi.org/10.1080/09298215.2024.2442361},
	doi = {10.1080/09298215.2024.2442361},
	number = {1-2},
	journal = {Journal of New Music Research},
	author = {Vigliensoni, Gabriel and and Fiebrink, Rebecca},
	month = mar,
	year = {2024},
	pages = {19--32},
}

@article{arzberger_reflexive_2024,
	title = {Reflexive {Data} {Curation}: {Opportunities} and {Challenges} for {Embracing} {Uncertainty} in {Human}–{AI} {Collaboration}},
	volume = {31},
	issn = {1073-0516},
	url = {https://doi.org/10.1145/3689042},
	doi = {10.1145/3689042},
	number = {6},
	journal = {ACM Trans. Comput.-Hum. Interact.},
	author = {Arzberger, Anne and Lupetti, Maria Luce and Giaccardi, Elisa},
	month = dec,
	year = {2024},
}

@inproceedings{garber_audiostellar_2020,
	title = {{AudioStellar}, an open source corpus-based musical instrument for latent sound structure discovery and sonic experimentation},
	url = {hdl.handle.net/2027/fulcrum.t435gg568},
	booktitle = {Proceedings of the 2021 {International} {Computer} {Music} {Conference}},
	author = {Garber, Leandro and Ciccola, Tomás and Amusategui, Juan Cruz},
	month = dec,
	year = {2020},
}

@article{shier_designing_2025,
	title = {Designing {Percussive} {Timbre} {Remappings}: {Negotiating} {Audio} {Representations} and {Evolving} {Parameter} {Spaces}},
	issn = {2220-4806},
	url = {http://nime.org/proceedings/2025/nime2025_66.pdf},
	doi = {10.5281/zenodo.15698926},
	journal = {Proceedings of the International Conference on New Interfaces for Musical Expression},
	author = {Shier, Jordie and Constanzo, Rodrigo and Saitis, Charalampos and Robertson, Andrew and McPherson, Andrew},
	editor = {Cavdir, Doga and Berthaut, Florent},
	month = jun,
	year = {2025},
	pages = {452--461},
}

@incollection{mudd_material-oriented_2019,
	address = {Cham},
	title = {Material-{Oriented} {Musical} {Interactions}},
	isbn = {978-3-319-92069-6},
	url = {https://doi.org/10.1007/978-3-319-92069-6_8},
	booktitle = {New {Directions} in {Music} and {Human}-{Computer} {Interaction}},
	publisher = {Springer International Publishing},
	author = {Mudd, Tom},
	editor = {Holland, Simon and Mudd, Tom and Wilkie-McKenna, Katie and McPherson, Andrew and Wanderley, Marcelo M.},
	year = {2019},
	doi = {10.1007/978-3-319-92069-6_8},
	pages = {123--133},
}

@inproceedings{schwarz_sound_2012,
	address = {Ann Arbor, Michigan},
	title = {The {Sound} {Space} as {Musical} {Instrument}: {Playing} {Corpus}-{Based} {Concatenative} {Synthesis}},
	url = {http://www.nime.org/proceedings/2012/nime2012_120.pdf},
	doi = {10.5281/zenodo.1180593},
	booktitle = {Proceedings of the {International} {Conference} on {New} {Interfaces} for {Musical} {Expression}},
	publisher = {University of Michigan},
	author = {Schwarz, Diemo},
	year = {2012},
}

@article{gaver_annotated_2012,
	title = {Annotated portfolios},
	volume = {19},
	issn = {1072-5520},
	url = {https://doi.org/10.1145/2212877.2212889},
	doi = {10.1145/2212877.2212889},
	number = {4},
	journal = {Interactions},
	author = {Gaver, Bill and Bowers, John},
	month = jul,
	year = {2012},
	pages = {40--49},
}

@article{zheng_exploring_2025,
	title = {Exploring {Gestural} {Affordances} in {Audio} {Latent} {Space} {Navigation}},
	volume = {7},
	issn = {2624-9898},
	url = {https://www.frontiersin.org/journals/computer-science/articles/10.3389/fcomp.2025.1575202/full},
	doi = {10.3389/fcomp.2025.1575202},
	language = {English},
	urldate = {2025-09-19},
	journal = {Frontiers in Computer Science},
	author = {Zheng, Shuoyang Jasper and Xambó Sedó, Anna and Bryan-Kinns, Nick},
	month = sep,
	year = {2025},
}

@misc{caillon_rave_2021,
	title = {{RAVE}: {A} variational autoencoder for fast and high-quality neural audio synthesis},
	doi = {10.48550/arXiv.2111.05011},
	author = {Caillon, Antoine and Esling, Philippe},
	year = {2021},
}

@inproceedings{tokui_latent_2025,
	address = {Daejeon, South Korea.},
	title = {Latent {Granular} {Resynthesis} using {Neural} {Audio} {Codecs}},
	doi = {10.48550/arXiv.2507.19202},
	booktitle = {Extended {Abstracts} for the {Late}-{Breaking} {Demo} {Session} of the 26th {Int}. {Society} for {Music} {Information} {Retrieval} {Conf}.},
	author = {Tokui, Nao and Baker, Tom},
	year = {2025},
}

@inproceedings{swift_panic_2025,
  author       = {Swift, Ben},
  title        = {PANIC!},
  booktitle    = {Proceedings of the International Conference on New
                   Interfaces for Musical Expression
                  },
  year         = 2025,
  pages        = {1--4},
  publisher    = {Zenodo},
  month        = jun,
  venue        = {Canberra, Australia},
  doi          = {10.5281/zenodo.17800975},
}

@article{nick_unstable_2025,
	title = {Unstable audio: code bending text-to-music generation},
	url = {https://aes2.org/publications/elibrary-page/?id=23004},
	number = {15},
	journal = {Journal of the Audio Engineering Society},
	author = {Collins, Nick},
	month = oct,
	year = {2025},
}

@inproceedings{welker_flowdec_2025,
	title = {{FlowDec}: {A} flow-based full-band general audio codec with high perceptual quality},
	url = {https://openreview.net/pdf?id=uxDFlPGRLX},
	booktitle = {The {Thirteenth} {International} {Conference} on {Learning} {Representations}},
	author = {Welker, Simon and Le, Matthew and Chen, Ricky T. Q. and Hsu, Wei-Ning and Gerkmann, Timo and Richard, Alexander and Wu, Yi-Chiao},
	year = {2025},
}

@inproceedings{shepardson_living_2023,
	address = {Mexico City, Mexico},
	title = {The {Living} {Looper}: {Rethinking} the {Musical} {Loop} as a {Machine} {Action}-{Perception} {Loop}},
	url = {http://nime.org/proceedings/2023/nime2023_32.pdf},
	doi = {10.5281/zenodo.11189164},
	booktitle = {Proceedings of the {International} {Conference} on {New} {Interfaces} for {Musical} {Expression}},
	author = {Shepardson, Victor and Magnusson, Thor},
	editor = {Ortiz, Miguel and Marquez-Borbon, Adnan},
	month = may,
	year = {2023},
	pages = {224--231},
}

@article{stefani_esteso_2024,
	title = {Esteso: {Interactive} {AI} {Music} {Duet} {Based} on {Player}-{Idiosyncratic} {Extended} {Double} {Bass} {Techniques}},
	issn = {2220-4806},
	url = {http://nime.org/proceedings/2024/nime2024_72.pdf},
	doi = {10.5281/zenodo.13904929},
	journal = {Proceedings of the International Conference on New Interfaces for Musical Expression},
	author = {Stefani, Domenico and Tomasetti, Matteo and Angeloni, Filiippo and Turchet, Luca},
	editor = {Bin, S. M. Astrid and Reed, Courtney N.},
	month = sep,
	year = {2024},
	pages = {490--498},
}

@inproceedings{meyer_interactive_2025,
	address = {New York, NY, USA},
	series = {C\&{C} '25},
	title = {Interactive {Movement}-to-{Audio} with {Pre}-{Trained} {Neural} {Networks}},
	isbn = {979-8-4007-1289-0},
	url = {https://doi.org/10.1145/3698061.3734415},
	doi = {10.1145/3698061.3734415},
	booktitle = {Proceedings of the 2025 {Conference} on {Creativity} and {Cognition}},
	publisher = {Association for Computing Machinery},
	author = {Meyer, Joseph and Bryan-Kinns, Nick and Fdili Alaoui, Sarah and Grierson, Mick and Fiebrink, Rebecca},
	year = {2025},
	pages = {491--493},
}

@inproceedings{khallaghi_squishysonics_2025,
	address = {New York, NY, USA},
	series = {{CHI} {EA} '25},
	title = {{SquishySonics}: {A} {Deformable} {Interface} for the {Physical} {Control} of {Real}-time {AI} {Sound} {Generation} {Tools}},
	isbn = {979-8-4007-1395-8},
	url = {https://doi.org/10.1145/3706599.3721175},
	doi = {10.1145/3706599.3721175},
	booktitle = {Proceedings of the {Extended} {Abstracts} of the {CHI} {Conference} on {Human} {Factors} in {Computing} {Systems}},
	publisher = {Association for Computing Machinery},
	author = {Khallaghi, Trisha and Bennett, Pete and Tanaka, Atau},
	year = {2025},
}

@article{liu_two_2025,
	title = {Two {Sonification} {Methods} for the {MindCube}},
	issn = {2220-4806},
	url = {http://nime.org/proceedings/2025/nime2025_74.pdf},
	doi = {10.5281/zenodo.15698944},
	journal = {Proceedings of the International Conference on New Interfaces for Musical Expression},
	author = {Liu, Fangzheng and Blanchard, Lancelot and Haddad, Don D. and Paradiso, Joseph},
	editor = {Cavdir, Doga and Berthaut, Florent},
	month = jun,
	year = {2025},
	pages = {511--515},
}

@article{hein_transsonic_2024,
	title = {Transsonic {\textbar} {Sonic} {Fluidity}},
	issn = {2220-4806},
	url = {http://nime.org/proceedings/2024/nime2024_music_16.pdf},
	doi = {10.5281/zenodo.15028017},
	journal = {Proceedings of the International Conference on New Interfaces for Musical Expression},
	author = {Hein, Nicola Leonard and Yip, Viola},
	editor = {Pardue, Laurel Smith and Dahlstedt, Palle},
	month = sep,
	year = {2024},
	pages = {57--61},
}

@article{azimi_live_2025,
	title = {Live {Improvisation} with {Fine}-{Tuned} {Generative} {AI}: {A} {Musical} {Metacreation} {Approach}},
	issn = {2220-4806},
	url = {http://nime.org/proceedings/2025/nime2025_54.pdf},
	doi = {10.5281/zenodo.15698902},
	journal = {Proceedings of the International Conference on New Interfaces for Musical Expression},
	author = {Azimi, Misagh and Zareei, Mo H.},
	editor = {Cavdir, Doga and Berthaut, Florent},
	month = jun,
	year = {2025},
	pages = {389--393},
}

@inproceedings{privato_magnetic_2023,
	address = {Mexico City, Mexico},
	title = {Magnetic {Interactions} as a {Somatosensory} {Interface}},
	url = {http://nime.org/proceedings/2023/nime2023_54.pdf},
	doi = {10.5281/zenodo.11189218},
	booktitle = {Proceedings of the {International} {Conference} on {New} {Interfaces} for {Musical} {Expression}},
	author = {Privato, Nicola and Magnusson, Thor and Einarsson, Einar Torfi},
	editor = {Ortiz, Miguel and Marquez-Borbon, Adnan},
	month = may,
	year = {2023},
	pages = {387--393},
}

@article{levitin_control_2002,
	title = {Control parameters for musical instruments: a foundation for new mappings of gesture to sound},
	volume = {7},
	doi = {10.1017/S135577180200208X},
	number = {2},
	journal = {Organised Sound},
	author = {Levitin, Daniel J. and McAdams, Stephen and Adams, Robert L.},
	year = {2002},
	pages = {171--189},
}

@book{barad_meeting_2007,
	title = {Meeting the {Universe} {Halfway}: {Quantum} {Physics} and the {Entanglement} of {Matter} and {Meaning}},
	isbn = {978-0-8223-3901-4},
	urldate = {2025-10-07},
	publisher = {Duke University Press},
	author = {Barad, Karen},
	year = {2007},
}

@article{stanley_compositional_2007,
	title = {Compositional pattern producing networks: {A} novel abstraction of development},
	volume = {8},
	issn = {1573-7632},
	url = {https://doi.org/10.1007/s10710-007-9028-8},
	doi = {10.1007/s10710-007-9028-8},
	number = {2},
	journal = {Genetic Programming and Evolvable Machines},
	author = {Stanley, Kenneth O.},
	month = jun,
	year = {2007},
	pages = {131--162},
}

@misc{rafii_musdb18-hq_2019,
	title = {{MUSDB18}-{HQ} - an uncompressed version of {MUSDB18}},
	url = {https://zenodo.org/records/3338373},
	doi = {10.5281/zenodo.3338373},
	urldate = {2025-10-19},
	publisher = {Zenodo},
	author = {Rafii, Zafar and Liutkus, Antoine and Stöter, Fabian-Robert and Mimilakis, Stylianos Ioannis and Bittner, Rachel},
	month = aug,
	year = {2019},
}

@misc{ostermann_aam_2022,
	title = {{AAM}: {Artificial} {Audio} {Multitracks} {Dataset}},
	shorttitle = {{AAM}},
	url = {https://zenodo.org/records/5794629},
	doi = {10.5281/zenodo.5794629},
	urldate = {2025-10-19},
	publisher = {Zenodo},
	author = {Ostermann, Fabian and Vatolkin, Igor},
	month = jun,
	year = {2022},
}

@misc{yamagishi_cstr_2019,
	title = {{CSTR} {VCTK} {Corpus}: {English} {Multi}-speaker {Corpus} for {CSTR} {Voice} {Cloning} {Toolkit} (version 0.92), [sound]},
	copyright = {Creative Commons Attribution 4.0 International Public License},
	shorttitle = {{CSTR} {VCTK} {Corpus}},
	url = {https://doi.org/10.7488/ds/2645},
	doi = {10.7488/ds/2645},
	urldate = {2025-10-19},
	author = {Yamagishi, Junichi and Veaux, Christophe and MacDonald, Kirsten},
	month = nov,
	year = {2019},
}

@article{ahn_eco-sonic_2025,
	title = {Eco-{Sonic} {Interfaces} for {Embodied} {AI} {Sound} {Exploration}},
	issn = {2220-4806},
	url = {http://nime.org/proceedings/2025/nime2025_1.pdf},
	doi = {10.5281/zenodo.15698772},
	journal = {Proceedings of the International Conference on New Interfaces for Musical Expression},
	author = {Ahn, Sabina Hyoju and Millett, Ryan and Park, Seyeon},
	editor = {Cavdir, Doga and Berthaut, Florent},
	month = jun,
	year = {2025},
	pages = {1--5},
}

@inproceedings{hershey_cnn_2017,
	title = {{CNN} architectures for large-scale audio classification},
	doi = {10.1109/ICASSP.2017.7952132},
	booktitle = {2017 {IEEE} {International} {Conference} on {Acoustics}, {Speech} and {Signal} {Processing} ({ICASSP})},
	author = {Hershey, Shawn and Chaudhuri, Sourish and Ellis, Daniel P. W. and Gemmeke, Jort F. and Jansen, Aren and Moore, R. Channing and Plakal, Manoj and Platt, Devin and Saurous, Rif A. and Seybold, Bryan and Slaney, Malcolm and Weiss, Ron J. and Wilson, Kevin},
	year = {2017},
	pages = {131--135},
}

@misc{noel-hirst_sampling_2025,
	title = {Sampling the {Latent} {Space}: {Exploring} the {Creative} {Potential} of {Generative} {AI} {Through} the {Lens} of {Sample}-{Based} {Music} {Making}},
	url = {https://doi.org/10.5281/zenodo.16946825},
	publisher = {Zenodo},
	author = {Noel-Hirst, Ashley and Saitis, Charalampos and Bryan-Kinns, Nick},
	month = sep,
	year = {2025},
}

@phdthesis{james_developing_2005,
	title = {Developing a flexible and expressive realtime polyphonic wave terrain synthesis instrument based on a visual and multidimensional methodology},
	url = {https://ro.ecu.edu.au/theses/107},
	school = {Edith Cowan University},
	author = {James, Stuart},
	month = jan,
	year = {2005},
}

@article{arisholm_dynamic_2004,
	title = {Dynamic coupling measurement for object-oriented software},
	volume = {30},
	issn = {1939-3520},
	url = {https://doi.org/10.1109/TSE.2004.41},
	doi = {10.1109/TSE.2004.41},
	number = {8},
	urldate = {2025-10-20},
	journal = {IEEE Transactions on Software Engineering},
	author = {Arisholm, Erik and Briand, Lionel C. and Foyen, Audun},
	year = {2004},
	pages = {491 -- 506},
}

@inproceedings{bergstra_algorithms_2011,
	title = {Algorithms for {Hyper}-{Parameter} {Optimization}},
	volume = {24},
	booktitle = {Advances in {Neural} {Information} {Processing} {Systems}},
	publisher = {Curran Associates, Inc.},
	author = {Bergstra, James and Bardenet, Rémi and Bengio, Yoshua and Kégl, Balázs},
	editor = {Shawe-Taylor, J. and Zemel, R. and Bartlett, P. and Pereira, F. and Weinberger, K. Q.},
	year = {2011},
}

@article{caspe_waveform_2025,
	title = {Waveform {Autoencoding} at the {Edge} of {Perceivable} {Latency}},
	issn = {2220-4806},
	url = {http://nime.org/proceedings/2025/nime2025_10.pdf},
	doi = {10.5281/zenodo.15699550},
	journal = {Proceedings of the International Conference on New Interfaces for Musical Expression},
	author = {Caspe, Franco and McPherson, Andrew and Sandler, Mark},
	editor = {Cavdir, Doga and Berthaut, Florent},
	month = jun,
	year = {2025},
	pages = {73--76},
}

@inproceedings{symons_perceptron_2024,
	address = {New York, NY, USA},
	series = {{AM} '24},
	title = {The {Perceptron}: {A} {Multi}-player {Entangled} {Instrument} based on {Interpretive} {Mapping} and {Intra}-action.},
	isbn = {979-8-4007-0968-5},
	url = {https://doi.org/10.1145/3678299.3678338},
	doi = {10.1145/3678299.3678338},
	booktitle = {Proceedings of the 19th {International} {Audio} {Mostly} {Conference}: {Explorations} in {Sonic} {Cultures}},
	publisher = {Association for Computing Machinery},
	author = {Symons, Steve},
	year = {2024},
	pages = {385--391},
}

@inproceedings{benjamin_machine_2021,
	address = {New York, NY, USA},
	series = {{CHI} '21},
	title = {Machine {Learning} {Uncertainty} as a {Design} {Material}: {A} {Post}-{Phenomenological} {Inquiry}},
	isbn = {978-1-4503-8096-6},
	url = {https://doi.org/10.1145/3411764.3445481},
	doi = {10.1145/3411764.3445481},
	booktitle = {Proceedings of the 2021 {CHI} {Conference} on {Human} {Factors} in {Computing} {Systems}},
	publisher = {Association for Computing Machinery},
	author = {Benjamin, Jesse Josua and Berger, Arne and Merrill, Nick and Pierce, James},
	year = {2021},
}

@incollection{wiberg_material-centered_2018,
	title = {Material-{Centered} {Interaction} {Design}},
	isbn = {978-0-262-34469-2},
	url = {https://doi.org/10.7551/mitpress/10427.003.0006},
	booktitle = {The {Materiality} of {Interaction}: {Notes} on the {Materials} of {Interaction} {Design}},
	publisher = {The MIT Press},
	author = {Wiberg, Mikael},
	month = feb,
	year = {2018},
}

@article{gioti_composing_2022,
	title = {Composing the {Assemblage}: {Probing} {Aesthetic} and {Technical} {Dimensions} of {Artistic} {Creation} with {Machine} {Learning}},
	volume = {46},
	issn = {0148-9267},
	shorttitle = {Composing the {Assemblage}},
	url = {https://doi.org/10.1162/comj_a_00658},
	doi = {10.1162/comj_a_00658},
	number = {4},
	urldate = {2025-10-29},
	journal = {Computer Music Journal},
	author = {Gioti, Artemi-Maria and Einbond, Aaron and Born, Georgina},
	month = dec,
	year = {2022},
	pages = {62--80},
}

@inproceedings{zheng_material_2022,
	address = {New York, NY, USA},
	series = {{DIS} '22},
	title = {Material {Matters}: {Exploring} {Materiality} in {Digital} {Musical} {Instruments} {Design}},
	isbn = {978-1-4503-9358-4},
	url = {https://doi.org/10.1145/3532106.3533523},
	doi = {10.1145/3532106.3533523},
	booktitle = {Proceedings of the 2022 {ACM} {Designing} {Interactive} {Systems} {Conference}},
	publisher = {Association for Computing Machinery},
	author = {Zheng, Jianing and Bryan-Kinns, Nick and McPherson, Andrew P.},
	year = {2022},
	pages = {976--986},
}

@inproceedings{shelby_generative_2024,
	address = {New York, NY, USA},
	series = {{CHI} '24},
	title = {Generative {AI} in {Creative} {Practice}: {ML}-{Artist} {Folk} {Theories} of {T2I} {Use}, {Harm}, and {Harm}-{Reduction}},
	isbn = {979-8-4007-0330-0},
	url = {https://doi.org/10.1145/3613904.3642461},
	doi = {10.1145/3613904.3642461},
	booktitle = {Proceedings of the 2024 {CHI} {Conference} on {Human} {Factors} in {Computing} {Systems}},
	publisher = {Association for Computing Machinery},
	author = {Shelby, Renee and Rismani, Shalaleh and Rostamzadeh, Negar},
	year = {2024},
}

@article{caramiaux_explorers_2022,
	title = {"{Explorers} of {Unknown} {Planets}": {Practices} and {Politics} of {Artificial} {Intelligence} in {Visual} {Arts}},
	volume = {6},
	url = {https://doi.org/10.1145/3555578},
	doi = {10.1145/3555578},
	number = {CSCW2},
	journal = {Proc. ACM Hum.-Comput. Interact.},
	author = {Caramiaux, Baptiste and Fdili Alaoui, Sarah},
	month = nov,
	year = {2022},
}

@article{visi_sophtar_2024,
	title = {The {Sophtar}: a networkable feedback string instrument with embedded machine learning},
	issn = {2220-4806},
	url = {http://nime.org/proceedings/2024/nime2024_22.pdf},
	doi = {10.5281/zenodo.13904810},
	journal = {Proceedings of the International Conference on New Interfaces for Musical Expression},
	author = {Visi, Federico},
	editor = {Bin, S. M. Astrid and Reed, Courtney N.},
	month = sep,
	year = {2024},
	pages = {142--148},
}

@inproceedings{shier_real-time_2024,
	address = {Utrecht, Netherlands},
	title = {Real-time {Timbre} {Remapping} with {Differentiable} {DSP}},
	url = {http://nime.org/proceedings/2024/nime2024_55.pdf},
	doi = {10.5281/zenodo.13904884},
	booktitle = {Proceedings of the {International} {Conference} on {New} {Interfaces} for {Musical} {Expression}},
	author = {Shier, Jordie and Saitis, Charalampos and Robertson, Andrew and McPherson, Andrew},
	editor = {Bin, S. M. Astrid and Reed, Courtney N.},
	month = sep,
	year = {2024},
	pages = {377--385},
}

@inproceedings{clarke_longevity_2025,
	address = {Canberra, Australia},
	title = {Longevity of {Deep} {Generative} {Models} in {NIME}: {Challenges} and {Practices} for {Reactivation}},
	url = {http://nime.org/proceedings/2025/nime2025_32.pdf},
	doi = {10.5281/zenodo.15735662},
	booktitle = {Proceedings of the {International} {Conference} on {New} {Interfaces} for {Musical} {Expression}},
	author = {Clarke, Isaac and Rí, Francesco Ardan Dal and Masu, Raul},
	editor = {Cavdir, Doga and Berthaut, Florent},
	month = jun,
	year = {2025},
	pages = {224--230},
}

@inproceedings{robson_thinking_2024,
	address = {New York, NY, USA},
	series = {{CHI} '24},
	title = {Thinking with {Sound}: {Exploring} the {Experience} of {Listening} to an {Ultrasonic} {Art} {Installation}},
	isbn = {979-8-4007-0330-0},
	url = {https://doi.org/10.1145/3613904.3642616},
	doi = {10.1145/3613904.3642616},
	booktitle = {Proceedings of the 2024 {CHI} {Conference} on {Human} {Factors} in {Computing} {Systems}},
	publisher = {Association for Computing Machinery},
	author = {Robson, Nicole and McPherson, Andrew and Bryan-Kinns, Nick},
	year = {2024},
}

@inproceedings{gregorio_augmenting_2019,
	address = {Porto Alegre, Brazil},
	title = {Augmenting {Parametric} {Synthesis} with {Learned} {Timbral} {Controllers}},
	url = {http://www.nime.org/proceedings/2019/nime2019_paper085.pdf},
	doi = {10.5281/zenodo.3673025},
	booktitle = {Proceedings of the {International} {Conference} on {New} {Interfaces} for {Musical} {Expression}},
	publisher = {UFRGS},
	author = {Gregorio, Jeff and Kim, Youngmoo},
	editor = {Queiroz, Marcelo and Sedó, Anna Xambó},
	month = jun,
	year = {2019},
	pages = {431--436},
}

@inproceedings{weber_parameterized_2019,
	address = {Porto Alegre, Brazil},
	title = {Parameterized {Melody} {Generation} with {Autoencoders} and {Temporally}-{Consistent} {Noise}},
	url = {http://www.nime.org/proceedings/2019/nime2019_paper035.pdf},
	doi = {10.5281/zenodo.3672914},
	booktitle = {Proceedings of the {International} {Conference} on {New} {Interfaces} for {Musical} {Expression}},
	publisher = {UFRGS},
	author = {Weber, Aline and Alegre, Lucas Nunes and Torresen, Jim and Silva, Bruno C. da},
	editor = {Queiroz, Marcelo and Sedó, Anna Xambó},
	month = jun,
	year = {2019},
	pages = {174--179},
}

@inproceedings{bowers_hybrid_2014,
	address = {London, United Kingdom},
	title = {Hybrid {Resonant} {Assemblages}: {Rethinking} {Instruments}, {Touch} and {Performance} in {New} {Interfaces} for {Musical} {Expression}},
	url = {http://www.nime.org/proceedings/2014/nime2014_438.pdf},
	doi = {10.5281/zenodo.1178718},
	booktitle = {Proceedings of the {International} {Conference} on {New} {Interfaces} for {Musical} {Expression}},
	publisher = {Goldsmiths, University of London},
	author = {Bowers, John and Haas, Annika},
	month = jun,
	year = {2014},
	pages = {7--12},
}

@article{tremblay_enabling_2021,
	title = {Enabling {Programmatic} {Data} {Mining} as {Musicking}: {The} {Fluid} {Corpus} {Manipulation} {Toolkit}},
	volume = {45},
	issn = {0148-9267},
	url = {https://doi.org/10.1162/comj_a_00600},
	doi = {10.1162/comj_a_00600},
	number = {2},
	journal = {Computer Music Journal},
	author = {Tremblay, Pierre Alexandre and Roma, Gerard and Green, Owen},
	month = jun,
	year = {2021},
	pages = {9--23},
}

@inproceedings{tahiroglu_ai-terity_2021,
	address = {Shanghai, China},
	title = {{AI}-terity 2.0: {An} {Autonomous} {NIME} {Featuring} {GANSpaceSynth} {Deep} {Learning} {Model}},
	url = {https://nime.pubpub.org/pub/9zu49nu5},
	doi = {10.21428/92fbeb44.3d0e9e12},
	booktitle = {Proceedings of the {International} {Conference} on {New} {Interfaces} for {Musical} {Expression}},
	author = {Tahiroğlu, Koray and Kastemaa, Miranda and Koli, Oskar},
	month = jun,
	year = {2021},
}

@inproceedings{engel_neural_2017,
	address = {Sydney, NSW, Australia},
	series = {{ICML}'17},
	title = {Neural audio synthesis of musical notes with {WaveNet} autoencoders},
	booktitle = {Proceedings of the 34th {International} {Conference} on {Machine} {Learning} - {Volume} 70},
	publisher = {JMLR.org},
	author = {Engel, Jesse and Resnick, Cinjon and Roberts, Adam and Dieleman, Sander and Norouzi, Mohammad and Eck, Douglas and Simonyan, Karen},
	year = {2017},
	pages = {1068--1077},
}

@inproceedings{mitcheltree_neutone_2025,
	title = {Neutone {SDK}: {An} {Open} {Source} {Framework} for {Neural} {Audio} {Processing}},
	doi = {10.48550/arXiv.2508.09126},
	booktitle = {{AES} {International} {Conference} on {Artificial} {Intelligence} and {Machine} {Learning} for {Audio}},
	author = {Mitcheltree, Christopher and Teleaga, Bogdan and Fyfe, Andrew and Masuda, Naotake and Schäfer, Matthias and Bradic, Alfie and Tokui, Nao},
	year = {2025},
}

@article{bacon_approaches_2025,
	title = {Approaches to notation for embodied engagement with a novel neural network-based musical instrument},
	volume = {Volume 7 - 2025},
	issn = {2624-9898},
	url = {https://www.frontiersin.org/journals/computer-science/articles/10.3389/fcomp.2025.1597806},
	doi = {10.3389/fcomp.2025.1597806},
	journal = {Frontiers in Computer Science},
	author = {Bacon, Benjamin Keith and Auri, Leonardo and McKee, Holly Ambrozic},
	year = {2025},
}

@inproceedings{kilgour_frechet_2019,
	title = {Fréchet {Audio} {Distance}: {A} {Reference}-{Free} {Metric} for {Evaluating} {Music} {Enhancement} {Algorithms}},
	doi = {10.21437/Interspeech.2019-2219},
	booktitle = {Interspeech 2019},
	author = {Kilgour, Kevin and Zuluaga, Mauricio and Roblek, Dominik and Sharifi, Matthew},
	year = {2019},
	pages = {2350--2354},
}

@inproceedings{kumar_high-fidelity_2023,
	address = {Red Hook, NY, USA},
	series = {{NIPS} '23},
	title = {High-fidelity audio compression with improved {RVQGAN}},
	booktitle = {Proceedings of the 37th {International} {Conference} on {Neural} {Information} {Processing} {Systems}},
	publisher = {Curran Associates Inc.},
	author = {Kumar, Rithesh and Seetharaman, Prem and Luebs, Alejandro and Kumar, Ishaan and Kumar, Kundan},
	year = {2023},
}

@inproceedings{akten_deep_2018,
	title = {Deep {Meditations}: {Controlled} navigation of latent space},
	url = {https://nips2018creativity.github.io/doc/Deep_Meditations.pdf},
	booktitle = {{NeurIPS} 2018 {Workshop} on {Machine} {Learning} for {Creativity} and {Design}},
	author = {Akten, Memo and Fiebrink, Rebecca and Grierson, Mick},
	month = dec,
	year = {2018},
}

@inproceedings{evans_stable_2025,
	title = {Stable {Audio} {Open}},
	doi = {10.1109/ICASSP49660.2025.10888461},
	booktitle = {{ICASSP} 2025 - 2025 {IEEE} {International} {Conference} on {Acoustics}, {Speech} and {Signal} {Processing} ({ICASSP})},
	author = {Evans, Zach and Parker, Julian D. and Carr, CJ and Zukowski, Zack and Taylor, Josiah and Pons, Jordi},
	year = {2025},
	pages = {1--5},
}

@inproceedings{rahimi_random_2007,
	series = {{NIPS}'07},
	title = {Random features for large-scale kernel machines},
	isbn = {978-1-60560-352-0},
	booktitle = {Proceedings of the 21st {International} {Conference} on {Neural} {Information} {Processing} {Systems}},
	publisher = {Curran Associates Inc.},
	author = {Rahimi, Ali and Recht, Benjamin},
	year = {2007},
	pages = {1177--1184},
}

@inproceedings{tralie_concatenator_2024,
	title = {The {Concatenator}: {A} {Bayesian} {Approach} to {Real} {Time} {Concatenative} {Musaicing}},
	doi = {10.5281/zenodo.14877471},
	booktitle = {Proceedings of the 25th {International} {Society} for {Music} {Information} {Retrieval} {Conference}},
	publisher = {ISMIR},
	author = {Tralie, Christopher J. and Cantil, Ben},
	month = nov,
	year = {2024},
	pages = {882--889},
}

@inproceedings{pasini_music2latent_2024,
	title = {{Music2Latent}: {Consistency} {Autoencoders} for {Latent} {Audio} {Compression}},
	doi = {10.5281/zenodo.14877289},
	booktitle = {Proceedings of the 25th {International} {Society} for {Music} {Information} {Retrieval} {Conference}},
	publisher = {ISMIR},
	author = {Pasini, Marco and Lattner, Stefan and Fazekas, George},
	month = nov,
	year = {2024},
	pages = {111--119},
}

@inproceedings{nistal_diff--riff_2024,
	title = {Diff-a-{Riff}: {Musical} {Accompaniment} {Co}-{Creation} via {Latent} {Diffusion} {Models}},
	doi = {10.5281/zenodo.14877327},
	booktitle = {Proceedings of the 25th {International} {Society} for {Music} {Information} {Retrieval} {Conference}},
	publisher = {ISMIR},
	author = {Nistal, Javier and Pasini, Marco and Aouameur, Cyran and Grachten, Maarten and Lattner, Stefan},
	month = nov,
	year = {2024},
	pages = {272--280},
}

@inproceedings{demerle_combining_2024,
	title = {Combining {Audio} {Control} and {Style} {Transfer} {Using} {Latent} {Diffusion}},
	doi = {10.5281/zenodo.14877437},
	booktitle = {Proceedings of the 25th {International} {Society} for {Music} {Information} {Retrieval} {Conference}},
	publisher = {ISMIR},
	author = {Demerlé, Nils and Esling, Philippe and Doras, Guillaume and Genova, David},
	month = nov,
	year = {2024},
	pages = {721--728},
}

@inproceedings{nabi_adding_2025,
	title = {Adding temporal musical controls on top of pretrained generative models},
	doi = {10.5281/zenodo.17811482},
	booktitle = {Proceedings of the 26th {International} {Society} for {Music} {Information} {Retrieval} {Conference}},
	publisher = {ISMIR},
	author = {Nabi, Sarah and Demerlé, Nils and Peeters, Geoffroy and Bevilacqua, Frederic and Esling, Philippe},
	month = sep,
	year = {2025},
	pages = {793--800},
}


\appendix

\section{Experiments Details}

We present full details for each task in Section~\ref{sec_exp}.

\subsection{Task Details}
\label{apx_exp}

Our initial effort was implemented on a 1D sound space. A 5-sec audio segment is arranged on a timeline, and the mapping model described in Section~\ref{sec_principles} is trained with time coordinates paired with the latent trajectory encoded from the audio segment. A timeline $V = \{v_1, v_2, \dots,v_t\} \in [0,1] ^ {1\times t}$ is used for training and an alternative timeline $V^\prime$ offset by $\phi \sim \mathcal{N}(0, 1e^{-6})$ for testing. 

Three audio datasets are used as sources for audio segments to cover a range of timbre variance: (i) all unmixed drum tracks in MUSDB18 \cite{rafii_musdb18-hq_2019}, (ii) all unmixed tracks in AAM \cite{ostermann_aam_2022} with instruments that are in the ``Bowed'' and ``Guitars'' category, and (iii) speech recording in VCTK \cite{yamagishi_cstr_2019}. For each audio segment in a collection, an MLP is trained to recreate the latent trajectory of that segment given the training timeline $V$. For each audio collection, a hyperparameter search \cite{bergstra_algorithms_2011} is performed to optimise network and training configuration. For M2L, SAO, and FlowDec, pre-trained model provided by the original works is used across all audio collections. For RAVE, three pretrained models from online repositories \cite{caillon_rave_2021, intelligent_instruments_lab_rave-models_2023} are used for three audio collections, respectively. This is due to the common practice of training RAVEs on domain-specific datasets.

\subsection{Additional Results}
\label{apx_results}

Standard MLPs without Fourier features are used as benchmarks to validate the benefit of applying Fourier feature mappings. Peak Signal-to-Noise Ratio (PSNR) is used to compare the regressed latent trajectories with the original latent trajectories. Table~\ref{tab_psnr_full} shows that integrating Fourier features offers better latent PSNRs. The better PSNR on RAVE is suspected to be due to its smaller latent dimensionality than the others. The better PSNR on M2L is suspected to be due to its larger compression ratio \cite{pasini_music2latent_2024}, so the same length audio segment is encoded to fewer latent vectors, resulting in less dense input coordinates.

Fréchet Audio Distance (FAD) \cite{kilgour_frechet_2019} is used to evaluate the overall similarity between original audio collections and reconstructed audio collections, which contain those reconstructed from the original latent trajectories (i.e., directly forward passed by encoder-decoder), and the regressed latent trajectories by MLPs. We use the VGGish \cite{hershey_cnn_2017} embeddings to compute the audio distance. Table~\ref{tab_fad_full} shows that with Fourier features mapping, the regression of latent trajectories only minimally degrades FAD compared to the encoder-decoder reconstruction, which largely depends on the capacity of the autoencoder itself.


\begin{table*}[t]
    \centering
    \caption{Regressing trajectories with the three audio collection across four autoencoders.}
    \label{tab_metrics_full}
    \begin{subtable}{\linewidth}
        \centering
        \small
        \caption{Peak Signal-to-Noise Ratio between original latent trajectories and regressed trajectories (higher is better).}
        \label{tab_psnr_full}
        \begin{tabular}{l|cccc|cccc|cccc}
            \toprule
            & \multicolumn{12}{c}{\textbf{Peak Signal-to-Noise Ratio $\uparrow$}} \\
            \cmidrule(lr){2-13}
            & \multicolumn{4}{c|}{\textbf{Drum}} & \multicolumn{4}{c|}{\textbf{String}} & \multicolumn{4}{c}{\textbf{VCTK}} \\
            \cmidrule(lr){2-13}
            Regression Config & RAVE & M2L & SAO & FlowDec & RAVE & M2L & SAO & FlowDec & RAVE & M2L & SAO & FlowDec \\
            \midrule
            No Fourier features & 27.516 & 29.774 & 21.697 & 23.530 & 19.463 & 29.003 & 20.666 & 22.837 & 18.345 & 29.075 & 19.459 & 20.768 \\
            Fourier features & \textbf{41.369} & \textbf{40.368} & \textbf{32.069} & \textbf{32.566} & \textbf{28.328} & \textbf{38.966} & \textbf{28.024} & \textbf{31.997} & \textbf{30.591} & \textbf{38.104} & \textbf{30.786} & \textbf{30.052} \\
            \bottomrule
        \end{tabular}
    \end{subtable}
    \\[0.75em]
    \begin{subtable}{\linewidth}
        \centering
        \small
        \caption{Fréchet Audio Distance of audio sets reconstructed from the original trajectories (encoder-decoder reconstruction) and two sets of regressed trajectories, comparing to the original audio samples (lower is better).}
        \label{tab_fad_full}
        \begin{tabular}{l|cccc|cccc|cccc}
            \toprule
            & \multicolumn{12}{c}{\textbf{Fréchet Audio Distance $\downarrow$}} \\
            \cmidrule(lr){2-13}
            & \multicolumn{4}{c|}{\textbf{Drum}} & \multicolumn{4}{c|}{\textbf{String}} & \multicolumn{4}{c}{\textbf{VCTK}} \\
            \cmidrule(lr){2-13}
            Regression Config & RAVE & M2L & SAO & FlowDec & RAVE & M2L & SAO & FlowDec & RAVE & M2L & SAO & FlowDec \\
            \midrule
            Encoder-decoder recon. & 7.664 & 6.659 & 2.324 & 2.009 & 2.718 & 2.859 & 2.642 & 2.102 & 3.313 & 5.439 & 1.546 & 1.039 \\
            \midrule
            No Fourier features & 14.188 & 12.354 & 28.163 & 28.069 & 17.665 & 9.267 & 10.542 & 10.322 & 25.801 & 8.451 & 50.474 & 19.592 \\
            Fourier features & \textbf{8.139} & \textbf{7.511} & \textbf{2.387} & \textbf{2.473} & \textbf{4.783} & \textbf{5.092} & \textbf{4.023} & \textbf{2.498} & \textbf{3.689} & \textbf{7.562} & \textbf{2.180} & \textbf{1.259} \\
            \bottomrule
        \end{tabular}
    \end{subtable}
\end{table*}

\subsection{Training loss frequency components}
\label{apx_freq}
\begin{figure}[ht]
  \centering
  \includegraphics[width=\linewidth]{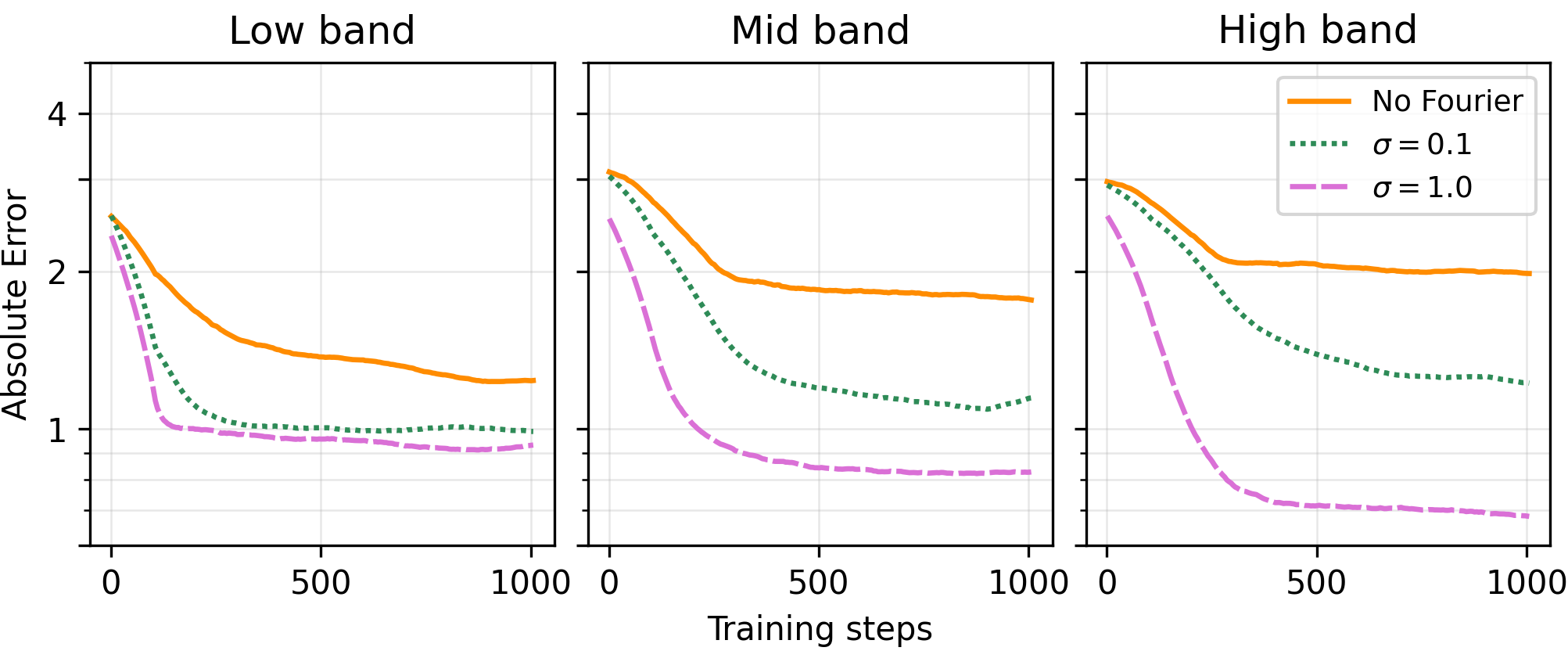}
  \caption{Absolute errors in each training step with respect to three frequency bands from low to high. }
  \Description{Absolute errors in each training step with respect to three frequency bands from low to high. }
  \label{fig_loss}
\end{figure}

We inspect the training error with respect to multiple frequency components to verify that the Fourier features are enabling the MLP to learn high-frequency details as claimed. We separate the training error with respect to different frequency components. Figure~\ref{fig_loss} shows a typical record of training errors: the model without Fourier features converges better in the low-frequency band but fails to converge further on the high-frequency band, whereas the two models that have Fourier features converge faster on all bands. This verifies that the introduction of Fourier features has allowed the model to resemble the high-dynamic nature of latent trajectories, and shows that the Gaussian scale $\sigma$ is a key configuration that affects the model behaviour, which we elaborate on in Section~\ref{sec_tuning}.

\subsection{On-the-fly Training and Inference}
\label{apx_fly}

\begin{figure*}[ht]
  \centering
  \includegraphics[width=0.8\linewidth]{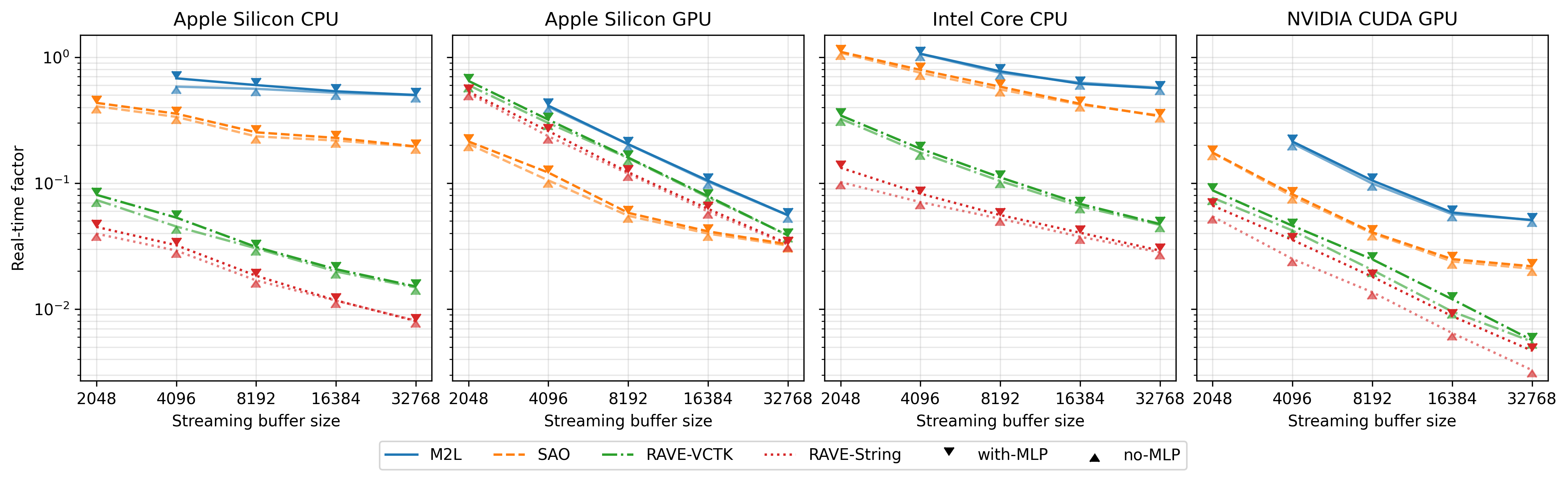}
  \caption{Mean real-time factor (lower the better) against buffer size, across all decoders and devices, comparing with and without the MLP. In all plots, mean values were computed over 100 runs per condition.}
  \Description{Mean real-time factor against buffer size, across all decoders and devices, comparing with and without the MLP. In all plots, mean values were computed over 100 runs per condition.}
  \label{fig_rtf_full}
\end{figure*}

We present full results for the on-the-fly training and inference evaluation from Section~\ref{sec_fly} in the main text. Results shown in Figure~\ref{fig_rtf_full} were tested on an Apple Silicon M4 Max MacBook Pro with 48GB RAM, and a Dell G16 7630 with NVIDIA RTX4070 GPU.

\section{Detailed Milestones and System Design}
\label{apx_milestones}

We present full details for each interface and milestone from Section~\ref{sec_situating} in the main text, in combination with major change logs and the details of the public engagement events shown in Table~\ref{tab_interface}.


\begin{figure*}[ht]
  \centering
  \includegraphics[width=0.8\linewidth]{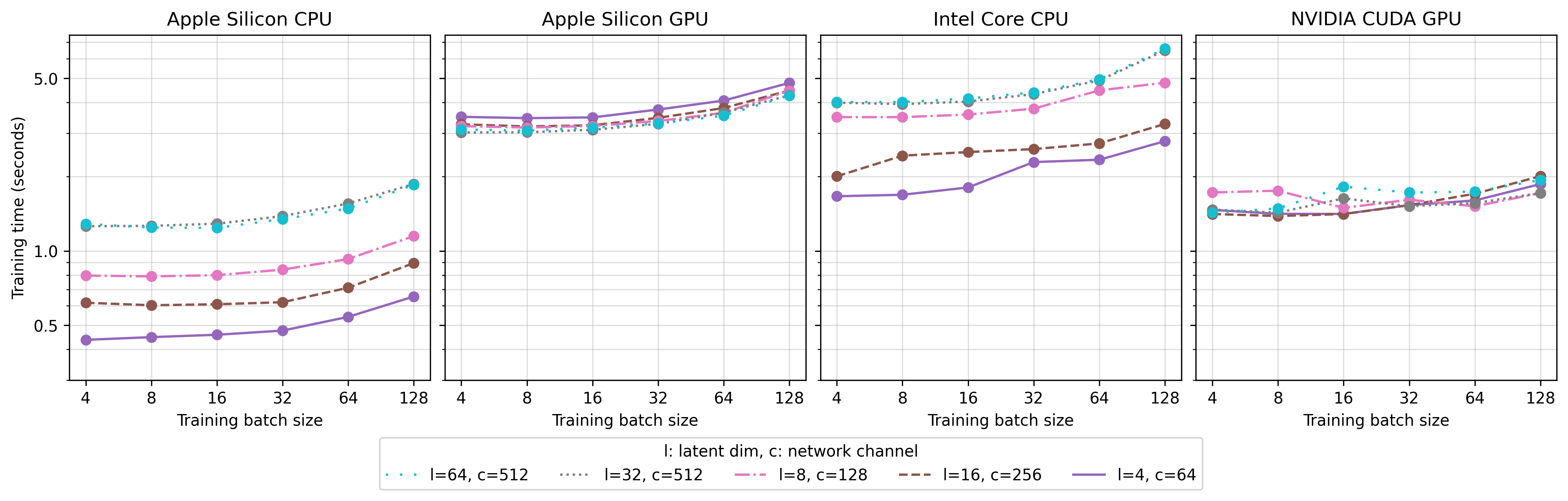}
  \caption{Average time needed to train the MLP for 1000 steps against training batch size, across various MLP configurations and devices. }
  \Description{Average time needed to train the MLP for 1000 steps against training batch size, across various MLP configurations and devices. }
  \label{fig_training_full}
\end{figure*}

\begin{table*}[ht]
\caption{Milestones of practice-led explorations that have guided the technical development.}
\begin{center}
\small
\begin{tabular}{l|l|p{38mm}|p{38mm}|p{34mm}|p{23mm}}
\toprule
\multirow{5}{*}{\STAB{\rotatebox[origin=c]{90}{\parbox[c]{1.2cm}{\centering \textbf{Change Logs}}}}}
 & \multicolumn{5}{p{100mm}}{\footnotesize\texttt{\textcolor{Maroon}{$++$} Exporting trained mapping model from PyTorch, loading it in MaxMSP}} \\
 & \multicolumn{5}{p{100mm}}{\footnotesize\texttt{\textcolor{Maroon}{$++$} Creating mapping visualisation in MaxMSP}} \\
 & \multicolumn{5}{p{100mm}}{\footnotesize\texttt{\textcolor{Maroon}{$++$} Tracking the stylus position as input}} \\
 & \multicolumn{5}{p{100mm}}{\footnotesize\texttt{\textcolor{Maroon}{$++$} Logging stylus behaviours}} \\
 & \multicolumn{5}{p{100mm}}{\footnotesize\texttt{\textcolor{Maroon}{$++$} Encapsulating nn$\sim$ for simplicity}} \\
\midrule
\multirow{3}{*}{\STAB{\rotatebox[origin=c]{90}{\parbox[c]{1.2cm}{\centering \textbf{Mile-stones}}}}}
 & \textbf{Time} & \textbf{Event} & \textbf{Type} & \textbf{Input Devices} & \textbf{Autoencoder} \\ 
\cmidrule{2-6}
 & Oct 2024 & Soundwalking Workshop & Interactive demo & Stylus and tablet interface & RAVE \\    
 & Dec 2024 & Interactive Digital Multimedia Techniques Concert & Composition and performance & Stylus and tablet interface & RAVE \\   
\midrule

\multirow{5}{*}{\STAB{\rotatebox[origin=c]{90}{\parbox[c]{1.2cm}{\centering \textbf{Change Logs}}}}}
 & \multicolumn{5}{p{160mm}}{\texttt{\footnotesize\textcolor{Maroon}{$++$} Defined independent objects nn.terrain$\sim$, nn.terrain.encode, and nn.terrain.gui for cohesion}} \\
 & \multicolumn{5}{p{160mm}}{\texttt{\footnotesize\textcolor{Maroon}{$++$} Interactive building sound space in MaxMSP: Encoding audio buffers to latents and interactive plotting spatial trajectories}} \\
 & \multicolumn{5}{p{100mm}}{\footnotesize\texttt{\textcolor{OliveGreen}{$--$} Removed nn$\sim$ to expose latent signals for appropriation}} \\
 & \multicolumn{5}{p{160mm}}{\texttt{\footnotesize\textcolor{Maroon}{$++$} Modulating stylus coordinates to program the trajectory playback}} \\
 & \multicolumn{5}{p{100mm}}{\texttt{\footnotesize\textcolor{OliveGreen}{$--$} Logging stylus behaviours}} \\
\midrule
\multirow{3}{*}{\STAB{\rotatebox[origin=c]{90}{\parbox[c]{1.2cm}{\centering \textbf{Mile-stones}}}}}
 & \textbf{Time} & \textbf{Event} & \textbf{Type} & \textbf{Input Devices} & \textbf{Autoencoder} \\ 
\cmidrule{2-6}
 & Mar 2025 & Queen Mary Music Festival & Composition and performance & Mixing desk knobs & RAVE \\    
 & Mar 2025 & IRCAM Forum Workshop & Demonstration & Stylus and tablet interface & RAVE \\   
\midrule

\multirow{4}{*}{\STAB{\rotatebox[origin=c]{90}{\parbox[c]{1.2cm}{\centering \textbf{Change Logs}}}}}
 & \multicolumn{5}{p{100mm}}{\texttt{\footnotesize\textcolor{Maroon}{$++$} Refined nn.terrain.gui for trajectory playback}} \\
 & \multicolumn{5}{p{160mm}}{\texttt{\footnotesize\textcolor{Maroon}{$++$} Added nn.terrain.record to record trajectory, to record trajectories from higher dimensional control space (e.g., 3D)}} \\
 & \multicolumn{5}{p{160mm}}{\texttt{\footnotesize\textcolor{Maroon}{$++$} Exposed more MLP and Fourier feature's hyperparameters}} \\
 & \multicolumn{5}{p{100mm}}{\texttt{\footnotesize\textcolor{Maroon}{$++$} Added the checkpoint method to save trained terrain}} \\
\midrule
\multicolumn{5}{p{100mm}}{\textbf{Initial release of the MaxMSP objects package}} \\
\midrule
\multirow{3}{*}{\STAB{\rotatebox[origin=c]{90}{\parbox[c]{0.6cm}{\centering \textbf{Mile-stones}}}}}
 & \textbf{Time} & \textbf{Event} & \textbf{Type} & \textbf{Input Devices} & \textbf{Autoencoder} \\ 
\cmidrule{2-6}  
 & Jun 2025 & Sónar+D Project Area & Interactive demo & Touchscreen & RAVE, Music2Latent \\   
\midrule

\multirow{3}{*}{\STAB{\rotatebox[origin=c]{90}{\parbox[c]{1.0cm}{\centering \textbf{Change Logs}}}}}
 & \multicolumn{5}{p{100mm}}{\texttt{\footnotesize\textcolor{Maroon}{$++$} Multi-threaded nn.terrain$\sim$ for on-the-fly machine learning}} \\
 & \multicolumn{5}{p{160mm}}{\texttt{\footnotesize\textcolor{Maroon}{$++$} Added support for stereo channels to support the autoencoder in Stable Audio Open}} \\
 & \multicolumn{5}{p{160mm}}{} \\

\bottomrule
\end{tabular}
\label{tab_interface}
\end{center}
\end{table*}

\subsection*{Interface A: Embodied Sound Space Navigation}

The first GUI design using \texttt{\anon[project.name]{nn.terrain$\sim$}} focuses on exploring the embodied movement-sound coupling in the sound space. It offers a drawing pad as the control space, in which the stylus $(x,y)$ location is used to navigate in the sound space. The GUI is used with a stylus as the input device to explore real-time sonic feedback in response to stylus movements. It renders the sound space using the visualisation method described in Section~\ref{sec_tuning}. The interface is presented in a series of ``\anon[Anonymous Workshops]{Soundwalking Workshops}'' in which musicians are tasked to explore the GUI as an instrument and create musical scores on the drawing pad. It was also used in a guest performance that premiered at the \anon[One Anonymous Course Title]{Interactive Digital Media Technology} concert at \anon[One Anonymous University Name]{Queen Mary University of London}. 

At this stage, it only marks a minimal viable prototype of using pre-built \anon[anonymous]{terrains} at the inference stage. Consequently, it relies heavily on programming the MLP mapping model in Python and then exporting it to TorchScript\footnote{\url{https://docs.pytorch.org/docs/2.5/jit.html}} to load in Max. This offline manner of integrating the sound space led to a long waiting time between programming it and actually seeing and hearing it. In this way, the time for musical exploration was spent mostly in navigating the sound space, annotating sonically interesting zones on the drawing pad, and experimenting with expressive ways of re-enacting sounds in these zones.

\subsection*{Interface B: Interactive Sound Space Construction}

The second development phase targeted an interactive talk and demo at the \anon[Anonymous Workshop Name 2025]{IRCAM Forum Workshop 2025}, and a guest performance that premiered at \anon[One Anonymous Music Festival 2025]{Queen Mary Music Festival 2025}. The API development in this phase focused on the interactive construction of \anon[anonymous]{terrains}. We attempted to match the workflow of building \anon[anonymous]{terrains} in Max/MSP with the workflow in Python. As a result, the cycle of training pairs gathering, model training, monitoring, and inference was distributed to three interdependent Max objects. We considered concepts of coupling and cohesion \cite{arisholm_dynamic_2004} in software engineering when defining the objects' API. The GUI design focuses on presenting the \anon[anonymous]{terrain} building workflow in a straightforward way. In particular, loading customised audio samples, plotting trajectories on the 2D plane, and iterative training while monitoring the \anon[anonymous]{terrain} visualisation.

We considered potential ways of programming the stylus input. Therefore, the drawing pad allows for modulating the stylus coordinates with an external source. In addition, the autoencoder that was previously encapsulated in the system was removed. In this way, \texttt{\anon[project.name]{nn.terrain$\sim$}} only fulfils the coordinate-to-latent mapping, and the autoencoder real-time inference relies on the existing \texttt{nn$\sim$}\footnote{https://github.com/acids-ircam/nn\_tilde} object from ACIDS-IRCAM, which has already been widely used by practitioners. Given this flexibility, in Interface B we created the right-most panel that allows modulating the stylus $(x,y)$ as a way of programming the trajectories in a \anon[anonymous]{terrain}, mimicking the ``oscillations and orbits'' control in \cite{roma_agent-based_2023}.

\subsection*{Interface C: Programming Trajectory Playback}

The third development phase targeted an interactive project exhibition at \anon[One Festival Name]{Sónar+D 2025}. We considered potential ways of using the \anon[anonymous]{terrains} other than the stylus. Therefore, the input (control space coordinates) and output (latent vectors) were exposed to the API to allow for appropriation for other modes of interaction. The initial project idea was to create a ``2D Granular Synth'' or a ``2D Sampler'', in which the standard audio sample arranged on a 1D timeline is replaced by a 2D sample canvas. This is practically impossible when sampling in the waveform domain because the sound playback cannot interpolate between spaces that are not filled with audio samples, unless using a discrete corpus-based sound space, as in concatenative synthesis \cite{garber_audiostellar_2020, tralie_concatenator_2024}. However, when sampling in the latent domain, the learned MLP mapping model can cover the entire 2D canvas with continuous coordinate input, and also generalise to spaces that are not initially arranged with samples, by interpolation.

To achieve this, the \textit{\anon[anonymous.gui]{nn.terrain.gui}} object was refined. As shown in Figure~\ref{fig_interfaces}C, the user first loads audio samples and plots trajectories in the ``train mode'' the same way as the previous interface. Then in the ``playback mode'' they define a new set of trajectories from the 2D canvas as ``latent samples''. Trajectories are triggered by the pattern in the step sequencer on the left. In addition, it integrates the M2L \cite{pasini_music2latent_2024} model as the underlying autoencoder. As a result, more MLP hyperparameters were exposed to the API to allow for hyperparameter tuning given the versatility of M2L.

\section{Artists Statements}
\label{apx_artists}

Here we include the bio of the four artist-researchers (group) who participanted in the practice-based exploration in Section~\ref{sec_artistic}.

\anon[Anonymised for review]{\textbf{Keigo Yoshida} is a Tokyo-based artist and scientist who integrates neuroscience and computer science to explore novel forms of artistic expression. His notable works include the A/V performances "Propagation" and "Mineral Neurons", both presented at Sónar+D 2025, and "liberated frequencies" (Installation and A/V performance). The latter work was supported by Flying Tokyo 2024, involved collaboration with METI and Rhizomatiks, and was also presented at IRCAM Forum Workshops 2025 Hors-les-Murs}.

\anon[Anonymised for review]{\textbf{Nico García-Peguinho} is an artist and researcher investigating machine listening and algorithmic composition. His work has been exhibited and performed at venues including Creative Coding Utrecht (live-coding AV performance), Manchester Histories Festival Opening Ceremony (8-channel installation), and MANTIS festival (multichannel fixed compositions). He is a first-year PhD student at the Centre for Digital Music (C4DM), Queen Mary University of London, and a member of the Sensing the Forest project.} \anon[Anonymised for review]{\textbf{Nikhil Bullock} is a London-based musician and researcher whose music is influenced by UK club culture and algorithmic composition. His practice explores generative processes as a tool for sonic exploration. He is currently pursuing an MSc in Sound and Music Computing at Queen Mary University of London.}

\anon[Anonymised for review]{\textbf{Jiatong Liu} is a Creative Technologist studying at the Creative Coding Institute, University of the Arts London. Their practice and research revolve around Machine Learning, Composition and Film - combining narrative design, spatial design and technology}.

\anon[Anonymised for review]{\textbf{Dan Hearn} is a creative technologist and musician working with interaction, sound, and machine learning. His work centres on building tools and interfaces that invite exploration rather than automation. He works across software, hardware, and machine learning to develop prototypes, instruments, and installations}.

\end{document}